\documentclass[twocolumn,showpacs,preprintnumbers,amsmath,amssymb]{revtex4}

\usepackage{graphicx}
\usepackage{dcolumn}
\usepackage{bm}

\begin{document}


\title{Spectral density analysis of time correlation functions in lattice
QCD using the maximum entropy method}

\author{H Rudolf Fiebig}
\affiliation{Physics Department, FIU, 11200 SW 8th Street, Miami, Florida 33199, USA}
\email{fiebig@fiu.edu}

\date{\today}

\begin{abstract}
We study various aspects of extracting spectral information 
from time correlation functions of lattice QCD by means of Bayesian inference
with an entropic prior, the maximum entropy method (MEM).
Correlator functions of a heavy-light meson-meson system serve as a
repository for lattice data with diverse statistical quality.
Attention is given to spectral mass density
functions, inferred from the data, and their dependence on the parameters of the MEM.
We propose to employ simulated annealing, or cooling, to solve the Bayesian
inference problem, and discuss practical issues of the approach.
\end{abstract}


\maketitle

\section{\label{sec:intro}Introduction}

Numerical simulations of quantum chromodynamics (QCD) on a Euclidean space-time lattice
provides access to mass spectra of hadronic systems through the
analysis of time correlation functions. In theory the latter are
linear combinations of exponential functions
\begin{equation}
C(t,t_0)=Z_1e^{-E_1(t-t_0)}+Z_2e^{-E_2(t-t_0)}+\ldots\,,
\label{exp2}\end{equation}
where the $E_n$ are the excitation energies of the system and the strength coefficients
\begin{equation}
Z_n=|\langle n|\hat{\Phi}(t_0)|0\rangle|^2
\label{Zp}\end{equation}
are matrix elements of some vacuum-subtracted operator
$\hat{\Phi}(t_0)=\Phi(t_0)-\langle 0|\Phi(t_0)| 0\rangle$
between the vacuum $|0\rangle$ and a ground or excited state $|n\rangle, n>0$.
In practice the exponential model (\ref{exp2}) is fitted to noisy numerical
simulation `data'. The statistical quality of simulation data rarely is good
enough for the two-exponential fit
(\ref{exp2}) to succeed. It is common practice to look at the large-$t$ behavior of the
correlation function $C(t,t_0)$ in a $t$-interval where it is dominated by only one
exponential, with the lowest energy, and then make a one-parameter fit to a plateau of the
effective-mass function $\mu_{\rm eff}(t,t_0)=-{\partial}\ln C(t,t_0)/{\partial t}$.
Possible discretizations are
\begin{subequations}
\begin{eqnarray}
\mu_{\rm eff,0}(t,t_0)&=&-\ln\left(\frac{C(t+1,t_0)}{C(t,t_0)}\right)
\simeq m_{\rm eff,0}\label{eff0}\\
\mu_{\rm eff,1}(t,t_0)&=&\frac{C(t+1,t_0)}{C(t,t_0)}\simeq e^{-m_{\rm eff,1}}\label{eff1}\\
\mu_{\rm eff,2}(t,t_0)&=&\frac{C(t+1,t_0)-C(t-1,t_0)}{2C(t,t_0)}\label{eff2}\\
& &\simeq -\sinh(m_{\rm eff,2})\nonumber\\
\mu_{\rm eff,3}^2(t,t_0)&=&\frac{C(t+1,t_0)+C(t-1,t_0)-2C(t,t_0)}{C(t,t_0)}\nonumber\\
& &\simeq 2(\cosh(m_{\rm eff,3})-1)\label{eff3}\,.
\end{eqnarray}
\end{subequations}
The expressions after the $\simeq$ are the values of $\mu_{\rm eff}$ for a
pure plateau of mass $m_{\rm eff}$.
The procedure implies the selection of consecutive	time slices $t=t_1\ldots t_2$
for which $\mu_{\rm eff}={\rm const}$, within errors, and an appropriate fit.
The selection of this, so-called, plateau is a matter of judgment.
A condition for reliable results is that the correlation function (\ref{exp2})
is dominated by just one exponential term, usually the ground state.
The latter can be enhanced by the use of smeared operators \cite{Alexandrou:1994ti} and fuzzy link
variables \cite{Alb87a}.
This analysis procedure discourages consideration of excited states. In fact
it will only produce reliably results if those are suppressed.
Workarounds involve diagonalization of a correlation matrix of several operators
or variational techniques \cite{Morningstar:1999rf}. Those however, still rely on plateau selection
without utilizing the information contained in the entire available time-slice range of a
correlation function.

As lattice simulations of QCD now aim at excited hadron states, $N^\ast$'s for
example \cite{Lee:1998cx,Gockeler:2001db,Sasaki:2001nf}, this situation is
unsatisfactory. Alternative methods
employing Bayesian inference \cite{Jar96} are a viable option.
The maximum entropy method (MEM), which involves a particular choice of the
Bayesian prior probability, falls in this class.
Bayesian statistics \cite{Box73} is a classic subject
with a vast range of applications.
However, application within the context of lattice QCD is relatively new
\cite{Nakahara:1999vy,Nakahara:1999bm,Asakawa:2000pv,Asakawa:2000tr,Lepage:2001ym}.

In this work we report on our experience using the MEM for extracting spectral mass
density functions $\rho(\omega)$ from lattice-generated time correlators
\begin{equation}
C(t,t_0)=\int d\omega\,\rho(\omega) e^{-\omega(t-t_0)}\,,
\label{Crho}\end{equation}
where a discrete set of time slices $t$ is understood. Discretization of the $\omega$-integral
with reasonably fine resolution leads to an ambiguous problem where the number
of parameters values $\rho(\omega)$ is (typically much) larger than the number of 
lattice data $C(t,t_0)$.
In the MEM an entropy term involving the spectral density is used
as a Bayesian prior to infer $\rho(\omega)$ from the data.

We here apply MEM analysis to sets of lattice correlation functions of a meson-meson system.
Those particular simulations are aimed at learning about mechanisms of hadronic interaction.
This will be discussed separately \cite{Fie02c}.
The lattice data generated within that
project involve local and nonlocal operators. They exhibit a wide range of statistical
quality from `very good' to `marginally acceptable'.

Our focus here is to utilize those data as a testing ground for Bayesian MEM analysis.
In contrast to other works we employ simulated annealing to the solution of
the Bayesian inference problem.
The main aim of this work is to explore the feasibility of this approach for extracting
masses from a lattice simulation using realistic lattice data, including excitations.
For the most part this translates into studying the sensitivity of the method to
to its native parameters.

\section{\label{sec:BayesCF}Bayesian Inference for Curve Fitting}

From a Bayesian point of view the spectral density function $\rho$ in (\ref{Crho}) 
is a random variable subject to a certain probability distribution functional
${\cal P}[\rho]$. Solution of the curve fitting problem consists in
finding the function $\rho$ which maximizes the conditional probability
${\cal P}[\rho\leftarrow C]$, the {\em posterior probability},
given a `measured' data set $C$.
Computation of $\rho$ is then based on Bayes' theorem \cite{Jar96}
\begin{equation}
{\cal P}[\rho\leftarrow C]\, {\cal P}[C]
={\cal P}[C\leftarrow \rho]\, {\cal P}[\rho]\,,
\label{BayesT}\end{equation}
also known as `detailed balance' in a different context.
The functional ${\cal P}[C]$, the {\em evidence}, gives the probability of
measuring a data set $C$. The conditional probability ${\cal P}[C\leftarrow \rho]$,
the {\em likelihood function}, determines the probability of measuring $C$
given a spectral function $\rho$. 
Finally ${\cal P}[\rho]$, the Bayesian {\em prior}, defines a constraint
on the spectral density function $\rho$. Its choice is a matter of judgment.
Ideally, the prior should reflect the physics known about the system,
for example an upper limit on the hadronic mass scale.
The posterior probability is the product of the likelihood function and the prior
${\cal P}[\rho\leftarrow C] =
{\cal P}[C\leftarrow \rho]\, {\cal P}[\rho]/{\cal P}[C]$,
where the {\em evidence} merely plays the role of a normalization constant \cite{Jar96}.
Indeed, the normalization condition
$\int[d\rho]{\cal P}[\rho\leftarrow C]=1$
applied to (\ref{BayesT}) gives
${\cal P}[C] = \int[d\rho]{\cal P}[C\leftarrow \rho]\, {\cal P}[\rho]$.
Thus, for a fixed $C$, we have
\begin{equation}
{\cal P}[\rho\leftarrow C]\propto{\cal P}[C\leftarrow \rho]\, {\cal P}[\rho]\,.
\label{BayesT3}\end{equation}
The curve fitting problem requires the product of the {\em likelihood function}
and the {\em prior} function.

\subsection{\label{sec:spectralD}Spectral density}

Our lattice data come from correlation functions
built from heavy-light meson-meson operators
\begin{equation}
\Phi_v=v_1\Phi_1+v_2\Phi_2\,,
\label{Phiv}\end{equation}
where $\Phi_1$ and $\Phi_2$ involve local and non-local meson-meson fields,
respectively, at relative distance $r$, and $v$ are some coefficients \cite{Fiebig:2001mr,Fiebig:2001nn}.
On a finite lattice the corresponding correlator
$C_v(t,t_0)=\langle\hat{\Phi}^\dagger_v(t)\hat{\Phi}_v(t_0)\rangle$,
where $\hat{\Phi}=\Phi-\langle\Phi\rangle$,
has a purely discrete spectrum
\begin{equation}
C_v(t,t_0)=\sum_{n\neq 0}
|\langle n|\Phi_v(t_0)|0\rangle|^2
e^{-\omega_n(t-t_0)}\,.
\label{Cvn} \end{equation}
Here $|n\rangle$ denotes a complete set of states with energies $\omega_n$,
some of which may be negative due to periodic lattice boundary conditions
and operator structure.
Our normalization conventions for forward and backward going
propagators are determined by defining
\begin{equation}
\exp_T(\omega,t)=\Theta(\omega)e^{-\omega t}
+\Theta(-\omega)e^{+\omega(T-t)}\,,
\label{expT}\end{equation}
where $0\leq t< T$, and $\Theta$ denotes the step function.
We then expect the lattice data to fit the following model
\begin{equation}
F(\rho_T|t,t_0)=\int_{-\infty}^{+\infty}d\omega\/
\rho_T(\omega)\exp_T(\omega,t-t_0)\,,
\label{Fc}\end{equation}
where $\rho_T(\omega)$ is a spectral density function,
defined for positive (forward) and negative (backward) frequencies.
The requirement that the model be exact, $F(\rho_T|t,t_0)=C_v(t,t_0)$, leads to
\begin{eqnarray}
\rho_T(\omega)&=&\sum_{n\neq 0}\delta(\omega-\omega_n)\,
|\langle n|\Phi_v(t_0)|0\rangle|^2\times\nonumber\\
& &[\Theta(\omega_n)+\Theta(-\omega_n)e^{-\omega_n T}]\,.\label{rhoT}
\end{eqnarray}
Thus a discrete sum over $\delta$-peaks is the theoretical form of
the spectral function. 
Our objective is to compute $\rho_T(\omega)$ from lattice data using
Bayesian inference.

\subsection{\label{sec:likely}Likelihood function}

Toward this end we proceed to construct the likelihood function.
The lattice data come in the form of an average over $N_U$ gauge configurations
\begin{equation}
C_v(t,t_0)=\frac{1}{N_U}\sum_{n=1}^{N_U}C_v(U_n|t,t_0)\,,
\label{CNU}\end{equation}
where $C_v(U_n|t,t_0)$ is the value of an operator, in this case
$\hat{\Phi}^\dagger_v(t)\hat{\Phi}_v(t_0)$,
in one gauge field configuration $U_n$.
Correlation function data on different time slices are stochastically
dependent. Their errors are described by the covariance matrix
\begin{eqnarray}
\Gamma_v(t_1,t_2)=\frac{1}{N_U}\sum_{n=1}^{N_U}
&&\left(\rule{0mm}{4mm}C_v(t_1,t_0)-C_v(U_n|t_1,t_0)\right)\times\nonumber\\
&&\left(\rule{0mm}{4mm}C_v(t_2,t_0)-C_v(U_n|t_2,t_0)\right).\label{Ecov}
\end{eqnarray}
The $\chi^2$-distance of the spectral model (\ref{Fc}) from the lattice data then is
\begin{eqnarray}
\chi^2=\sum_{t_1,t_2}
&&\left(\rule{0mm}{4mm}C_v(t_1,t_0)-F(\rho_T|t_1,t_0)\right)\Gamma_v^{-1}(t_1,t_2) 
\times\nonumber\\
&&\left(\rule{0mm}{4mm}C_v(t_2,t_0)-F(\rho_T|t_2,t_0)\right).\label{Chi2}
\end{eqnarray}
For numerical work a discretization scheme of the $\omega$-integral in (\ref{Fc})
is required. Our choice is
\begin{equation}
F(\rho_T|t,t_0)\simeq\sum_{k=K_-}^{K_+}
\rho_k\,\exp_T(\omega_k,t-t_0)
\label{Fd}\end{equation}
where $\omega_k=\Delta\omega k$, $\Delta\omega$ is an appropriate (small) interval,
$\rho_k=\Delta\omega \rho_T(\omega_k)$, and $K_- < 0 < K_+$.

The likelihood function ${\cal P}[C\leftarrow \rho]$ describes the probability
distribution of the data $C$ given a certain parameter set $\rho$. If we imagine
that the data are obtained by a large number of measurements, at fixed $\rho$,
then the probability distribution for $C$ is Gaussian by virtue of the
central limit theorem,
\begin{equation}
{\cal P}[C\leftarrow \rho]\propto e^{-\chi^2/2}\,.
\label{Pchi}\end{equation}
This is the standard argument for employing the above form of the likelihood
function in the context of Bayesian inference \cite{Jar96,Bra76}.

\subsection{\label{sec:prior}Entropic prior}

In case some information is available about the physics of the system it can be used
to constrain the parameter space of the model. This is the role of the Bayesian prior.
In the standard approach plateau methods are a severe form of imposing restrictions.
A two-exponential fit (\ref{exp2}), if feasible, is less constraining. 
In a Bayesian context it is possible to gradually increase the number of
exponentials until convergence is reached. This is a strategy advocated
in \cite{Lepage:2001ym}, see also \cite{Morningstar:2001je}.
There, the model for the correlation function is $\sum_n A_n e^{-E_nt}$, initially
with small number of terms, which is then constrained by the Bayesian prior
$e^{-\sum_n[(A_n-\bar{A}_n)^2/2\bar{\sigma}_{A_n}^2
           +(E_n-\bar{E}_n)^2/2\bar{\sigma}_{E_n}^2]}$.
The quantities $\bar{A}_n,\bar{\sigma}_{A_n},\bar{E}_n,\bar{\sigma}_{E_n}$ are
input. Their choice is inspired by prior knowledge about the physics
of the system.

On the other hand, there is usually no {\em a priory} information about the
location and the strengths of the peaks in the mass spectrum.
The view that only {\em minimal information} is available about the spectral
density function can also be implemented in the Bayesian prior.
The information content, in the sense of \cite{Sha49,Jay57a,Jay57b}, is measured
by the entropy $S=-\sum_k \rho_k\ln(\rho_k/m)$, on some scale $m$.
Rather, a commonly used variant is the Shannon-Jaynes entropy \cite{Jar96}
\begin{equation}
{\cal S}[\rho]=\sum_{k=K_-}^{K_+}\left(\rho_k-m_k-\rho_k\ln\frac{\rho_k}{m_k}\right)\,.
\label{Smem}\end{equation}
Note that $\rho_k\ge 0$, according to (\ref{rhoT}).
The configuration $m=\{m_k : K_- \leq k \leq K_+\}$ is called the default model.
We have ${\cal S}\leq 0$, $\forall\rho$, while ${\cal S}=0 \iff \rho=m$.
The default model is a unique absolute maximum of ${\cal S}$.
Choosing the prior probability as
\begin{equation}
{\cal P}[\rho]\propto e^{\alpha{\cal S}}
\label{Pent}\end{equation}
entails that ${\cal P}[\rho]$ is maximal in the absence of information
about $\rho$. An argument for (\ref{Pent}) can be found in \cite{Jar96}.
The entropy strength $\alpha$ and the default model $m$ are parameters.

\subsection{\label{sec:compute}Computing the spectral density}

With (\ref{Pchi}) and (\ref{Pent}) the posterior probability (\ref{BayesT3}) becomes
\begin{equation}
{\cal P}[\rho\leftarrow C]\propto e^{-(\chi^2/2-\alpha{\cal S})}\,.
\label{Ppost}\end{equation}
We wish to maximize ${\cal P}[\rho\leftarrow C]$ with respect to $\rho$, at fixed $C$.
It can be shown that both $\chi^2[\rho]$ and $-{\cal S}[\rho]$ are convex functions of
$\rho=\{\rho_k : K_- \leq k \leq K_+\}$. Thus
\begin{equation}
W[\rho]=\chi^2/2-\alpha S
\label{Wrho}\end{equation}
has a unique absolute minimum.
The functional $W[\rho\/]$ is nonlinear and maximally nonlocal since all degrees
of freedom $\rho_k$ are coupled via the covariance matrix (\ref{Ecov}) in (\ref{Chi2}).
To find the minimum of $W[\rho]$ one option is to use singular value decomposition (SVD),
see \cite{Asakawa:2000tr}.

In keeping with the Bayesian probabilistic interpretation of $\rho$
an attractive alternative is to
employ stochastic methods to solve the optimization
problem $W[\rho\/]=\min$.
In this work we employ simulated annealing \cite{Kir84}, equivalently known
as cooling. The algorithm is based on the partition function
\begin{equation}
Z_W=\int [d\rho\/] e^{-\beta_W W[\rho\/]}\,.
\label{Zmem}\end{equation}
It involves the generation of equilibrium configurations $\rho$ while
gradually increasing $\beta_W$ from an initially small value, following
some annealing schedule. The latter is subject to experimentation. We have
used the power law
\begin{equation}
\beta_W(n)=(\beta_1-\beta_0)\left(n/N\right)^\gamma+\beta_0
\label{powerlaw}\end{equation}
with annealing steps $n=0\ldots N$ between an initial $\beta_0$ and
a final $\beta_1$. 

A standard Metropolis algorithm was used to generate configurations $\rho$
with the distribution in (\ref{Zmem}).
In consecutive sweeps local updates were done by multiplying the spectral parameters
with positive random numbers,  $\rho_k\rightarrow x\rho_k$.
Some experimenting showed that $\Gamma$-distributed random deviates of order two,
$p_a(x)=x^{a-1}e^{-x}/\Gamma(a), a=2$, work quite efficiently at an acceptance
rate centered at about 50\%.

\section{\label{sec:results}Results}

All simulations were done on an $L^3\times T=10^3\times 30$ lattice.
The gauge field and fermion actions are both anisotropic,
with bare aspect ratio of $a_s/a_t=3$, and tadpole improved.
The gauge field action is that of \cite{Morningstar:1999rf} with $\beta=2.4$,
leading to a spatial lattice constant of
$a_s\simeq 0.25{\rm fm}, a_s^{-1}\simeq 800{\rm MeV}$.
For the light fermions we use a clover improved Wilson action.
The hopping parameter $\kappa=0.0679$ results in a mass
ratio $m_\pi/m_\rho \simeq 0.75$.
Following \cite{Morningstar:1999rf} only spatial directions are
improved with spatial tadpole renormalization factors
$u_s=\langle\, \framebox(5,5)[t]{}\,\rangle^{1/4}$, while $u_t=1$
in the time direction.
Clover terms involving time directions are omitted.

Some guidance for a reasonable $\omega$-discretization (\ref{Fd}), of (\ref{Fc}),
may be derived from the physical value of the lattice constant $a_t$, and the
time extent $Ta_t$ of the lattice. Admissible lattice energies thus lie
approximately between $\pi/a_t\approx 7.5{\rm GeV}$ and $\pi/Ta_t\approx 250{\rm MeV}$,
or $\approx 3$ and $\approx 0.1$ in units of $a_t^{-1}$.
In practice these are somewhat extreme bounds.
Typical hadronic excitation energies are much less than $\pi/a_t\approx 7.5{\rm GeV}$.
The lower bound, on the other hand, may well be ignored as a criterion for choosing
the discretization interval $\Delta\omega$,
because the theoretical form of $\rho$ is a superposition of $\delta$-peaks.
Thus the resolution $\Delta\omega$ should be small,
in fact much smaller than $\approx 0.1$. A reasonable lower bound is the likely
statistical error on spectral masses.
For most of the results presented here
$\Delta\omega=0.04$, and $ K_-=-40, K_+=+80$,
leading to $-1.6\leq\omega\leq +3.2$, were used with (\ref{Fd}).

With the annealing schedule (\ref{powerlaw}),
we have used $N=2048$ cooling steps, at 128 sweeps per temperature,
starting at 
$\beta_0=1.0\times 10^{-5}\beta$ and ending at
$\beta_1=1.0\times 10^{+5}\beta$, with a geometric average of $\beta=1.0\times 10^{+3}$.
These choices are an outcome of experimentation.
With $\gamma\simeq 16.61$ in (\ref{powerlaw})
about half of the cooling steps operate in the regions
$\beta_W(n)<\beta$ and $\beta_W(n)>\beta$, respectively.
The average value $\beta$ is such that $\beta_W W[\rho\/]$
fluctuates about one at around $N/2$ cooling steps.
With the final annealing temperature kept constant, $\beta_W=\beta_1$, an
additional 1024 steps were done keeping 16 configurations $\rho$ 
in order to measure cooling fluctuations.

Results are robust within reasonable changes of the annealing schedule
parameters, they were used throughout this work.

\subsection{\label{sec:alpha}Entropy weight dependence}

The extent to which the spectral density $\rho$ depends
on the value of the entropy weight parameter $\alpha$, in (\ref{Wrho}),
is a primary concern.
We are interested in testing the $\alpha$ dependence for a case where both ground and
excited states are prominently present in a time correlation function.
For this reason we have constructed a mock correlator $C_{\rm X}(t,t_0)$.
Its building blocks were the eigenvalues of the $2\times 2$ correlation matrix
$C_{ij}(t,t_0) = \langle\hat{\Phi}_i^\dagger(t)\hat{\Phi}_j(t_0)\rangle$
using the above mentioned local and non-local meson fields.
Pieces of those were arbitrarily matched and enhanced in order to exhibit
a multi-exponential correlation function.
While $C_{\rm X}(t,t_0)$ bears no physical significance,
its rich structure provides a useful laboratory for testing the $\alpha$ dependence
of the spectral density function.

\begin{figure*}
\includegraphics[angle=90,width=84mm]{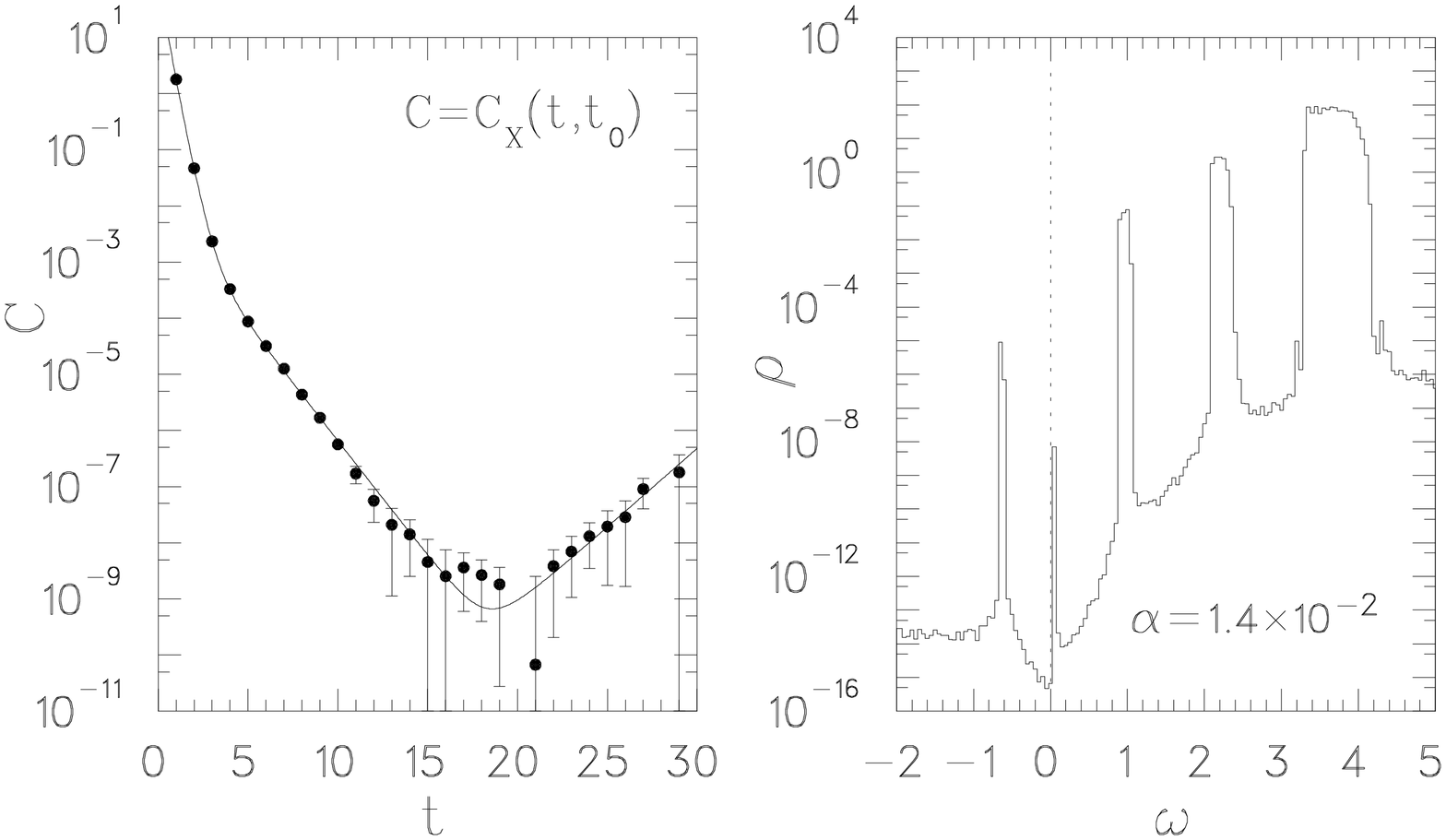}\hfill
\includegraphics[angle=90,width=84mm]{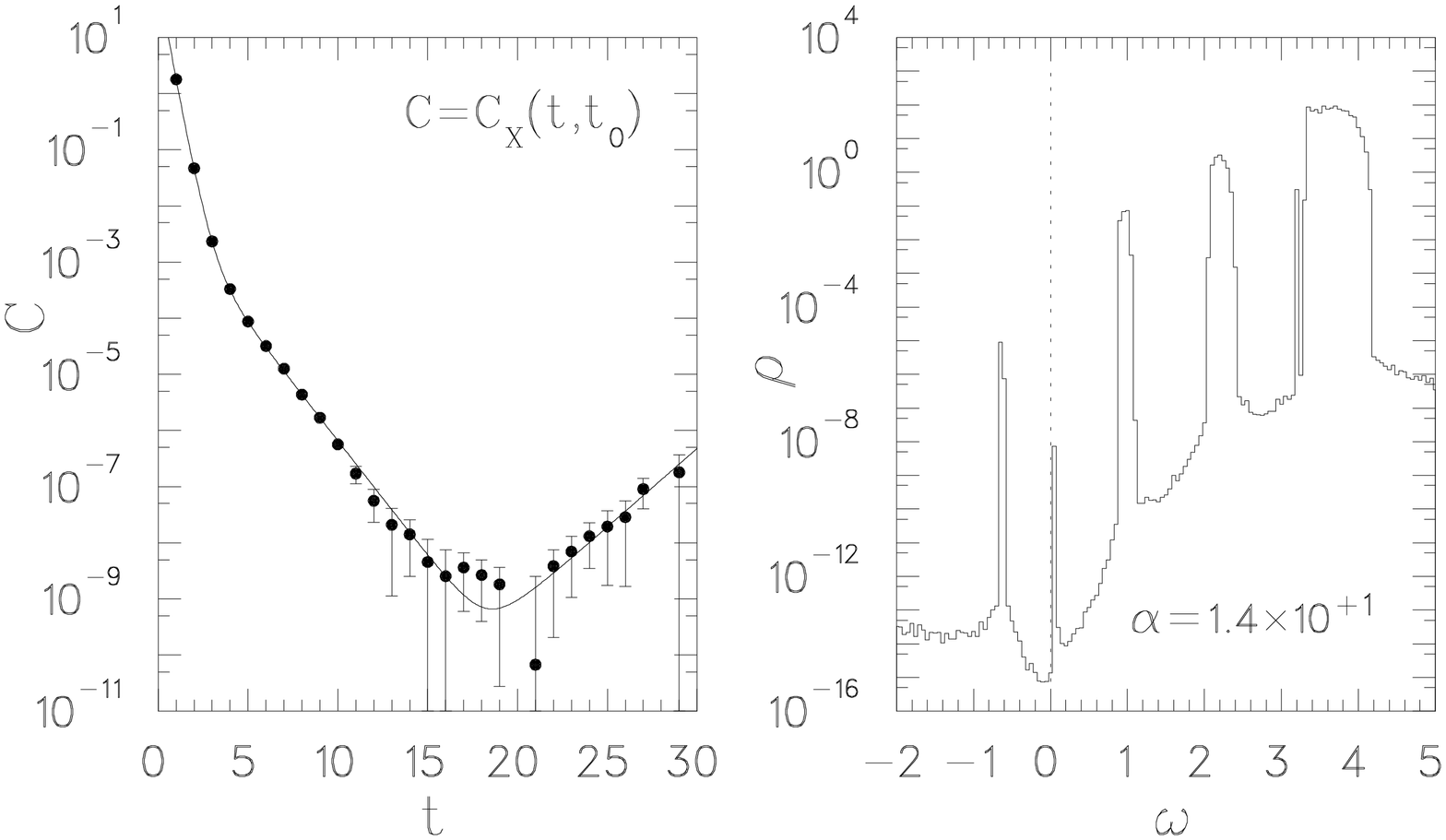}\\
\includegraphics[angle=90,width=84mm]{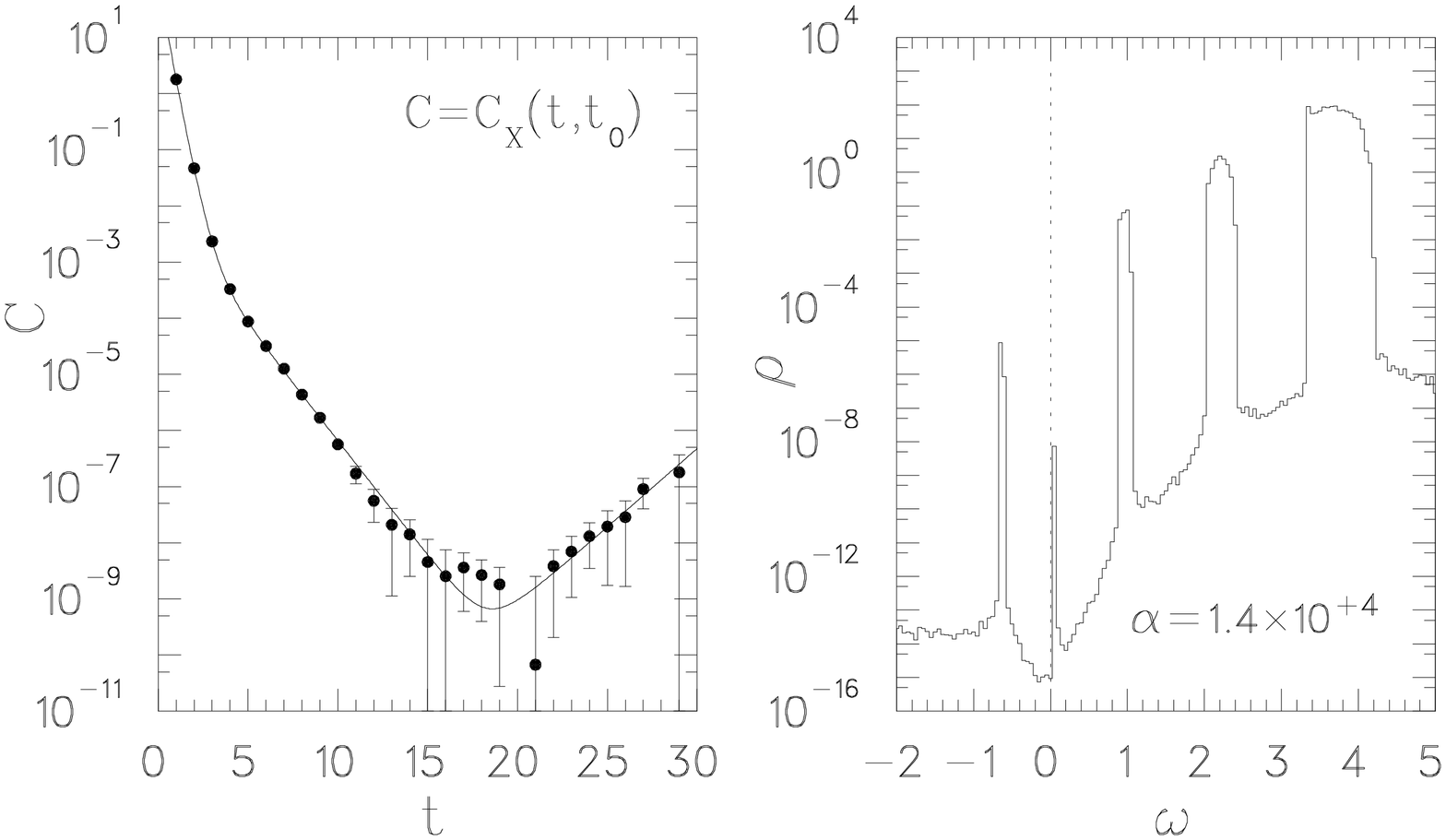}\hfill
\includegraphics[angle=90,width=84mm]{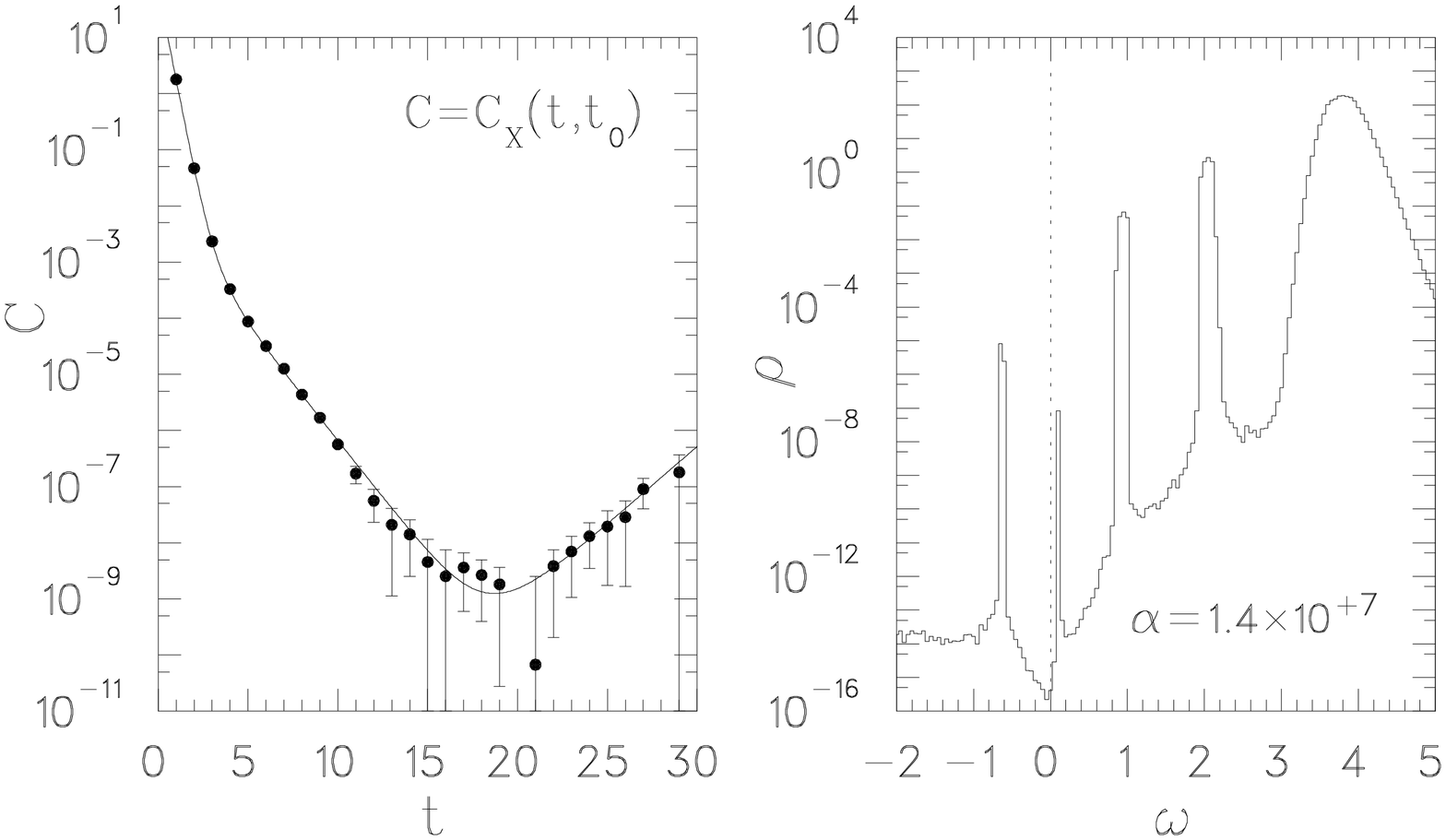}\\
\includegraphics[angle=90,width=84mm]{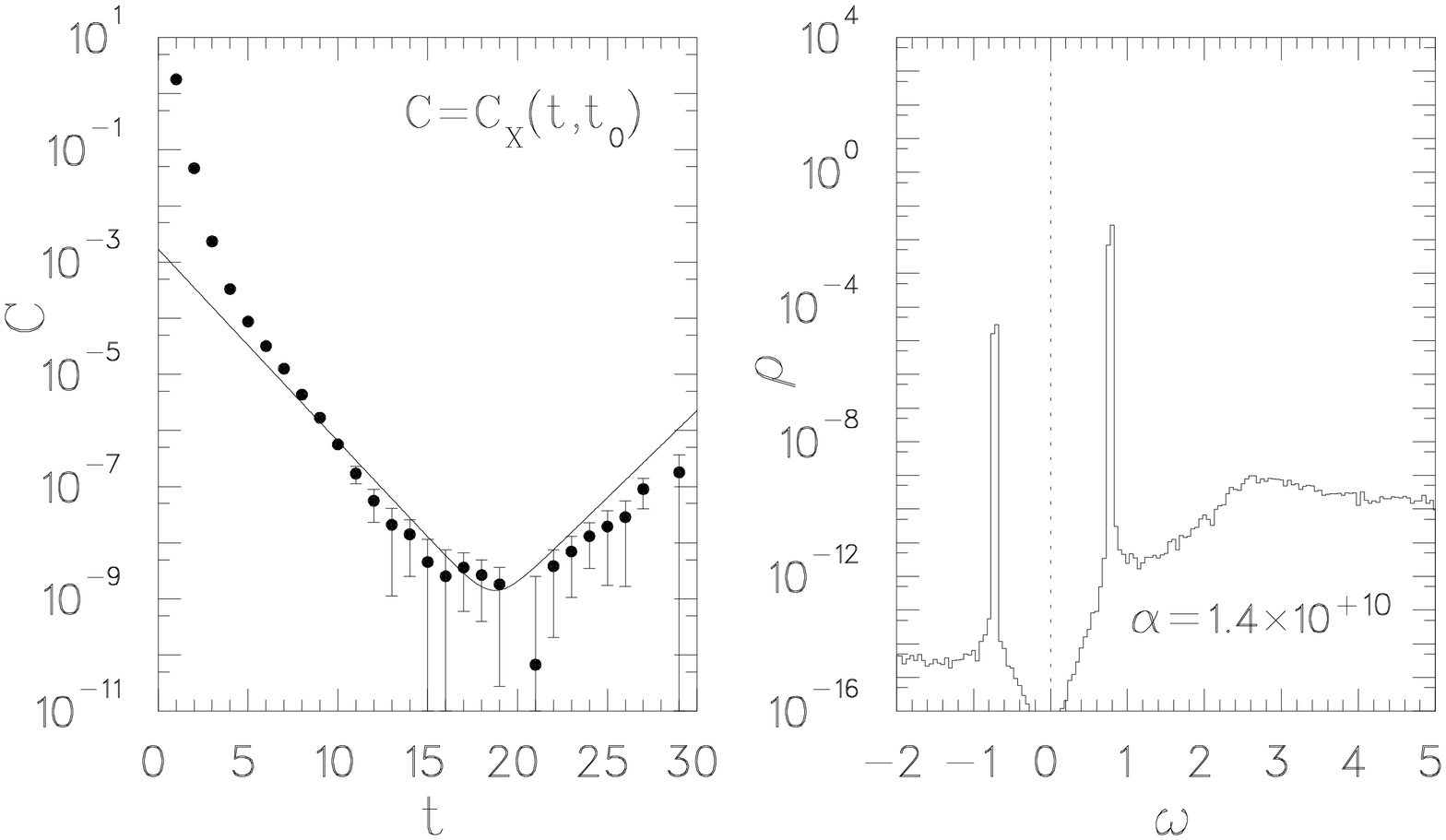}\hfill
\includegraphics[angle=90,width=84mm]{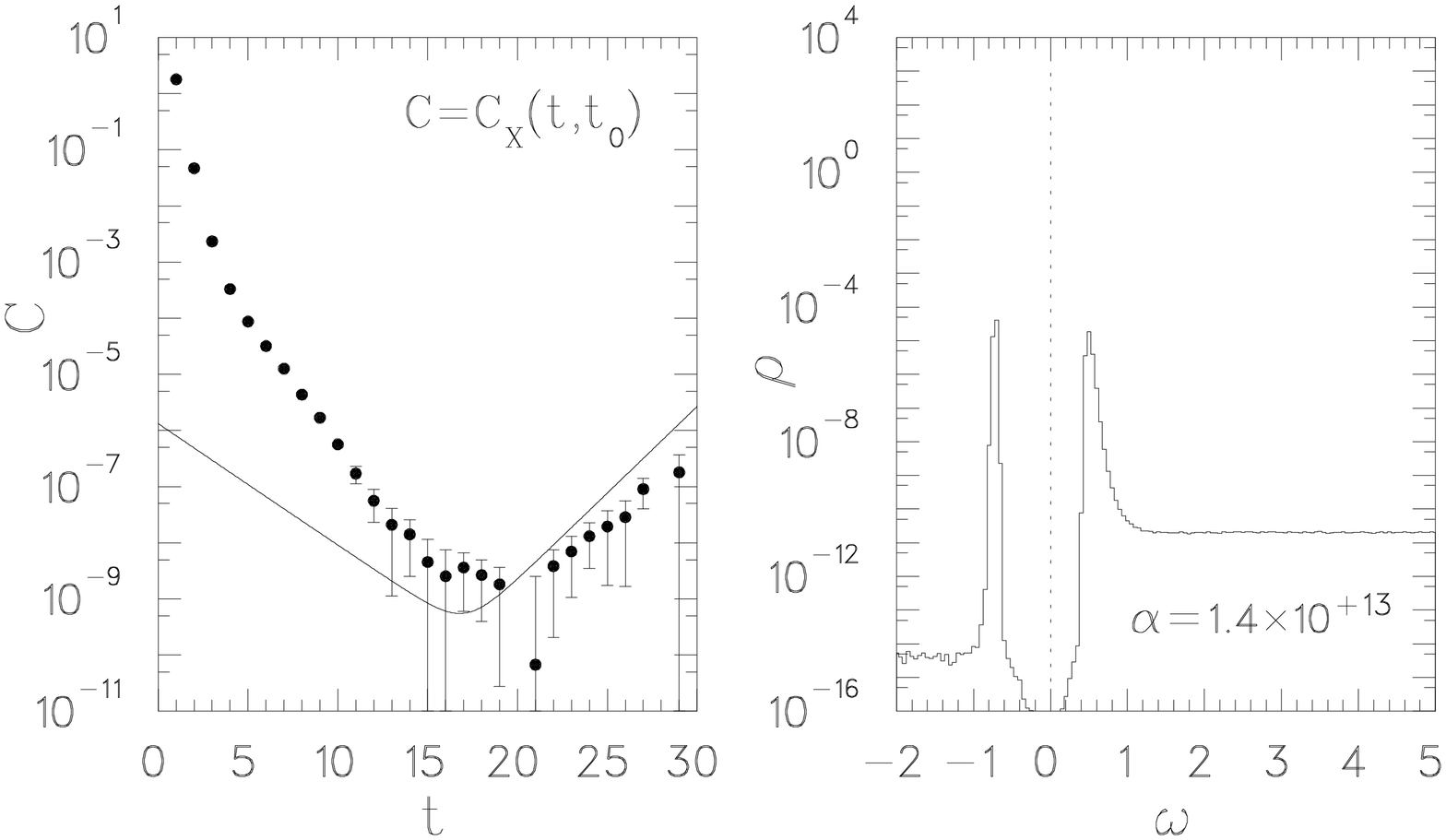}
\caption{\label{fig1}Mock time correlation functions $C_X$
shown with their Bayesian fits (solid lines), and the corresponding spectral densities
$\rho$. The sequence of six pairs of figures shows how the spectral fit evolves
through a change of the entropy weight $\alpha$ through 15 orders of magnitude.}
\end{figure*}

In Fig.~\ref{fig1} we show a sequence of six pairs of Bayesian fits to the
mock correlator $C_{\rm X}(t,t_0)$ and the corresponding spectral densities $\rho$
for a wide range of entropy weights $\alpha$. The stability of the global structure
of $\rho$ while $\alpha$ changes from $1.4\times 10^{-2}$ to $1.4\times 10^{+7}$ is
most notable\footnote{In \protect\cite{Fiebig:2001nn} and \protect\cite{Fiebig:2001mr}
a different
normalization of the covariance matrix (\protect\ref{Ecov}) was used. This can be
accounted for by a rescaling of the entropy weight $\alpha=(N_U-1)\bar{\alpha}$,
where $\bar{\alpha}$ refers to the above references and $N_U=708$.}.
As $\alpha$ becomes larger entire peaks vanish starting with the smallest one.
The reason is that the annealing action (\ref{Wrho}) gradually loses memory of
the data, contained in $\chi^2$, in favor of the entropy.
The fit at $\alpha=1.4\times 10^{+7}$ exhibits the onset of a smoothing of the micro
structure, starting with the largest peak.
This is the signature of emerging entropy dominance over the data. 
In practice this situation should be avoided. 
In our case entropy strengths in the region $\alpha < 10^{+6}$
over eight orders of magnitude give stable consistent results. 
It has been proposed that spectral functions be integrated over $\alpha$ to avoid
the parameter dependence \cite{Jar96}. Inspection of our results clearly indicates
that averaging over $\alpha$ would be without consequence to the gross structure
of $\rho$, only the micro structure would be affected.
Even the region $\alpha>10^{+6}$ could be included, since the magnitude of $\rho$
quickly becomes insignificant.

\begin{figure}
\includegraphics[angle=0,width=42mm]{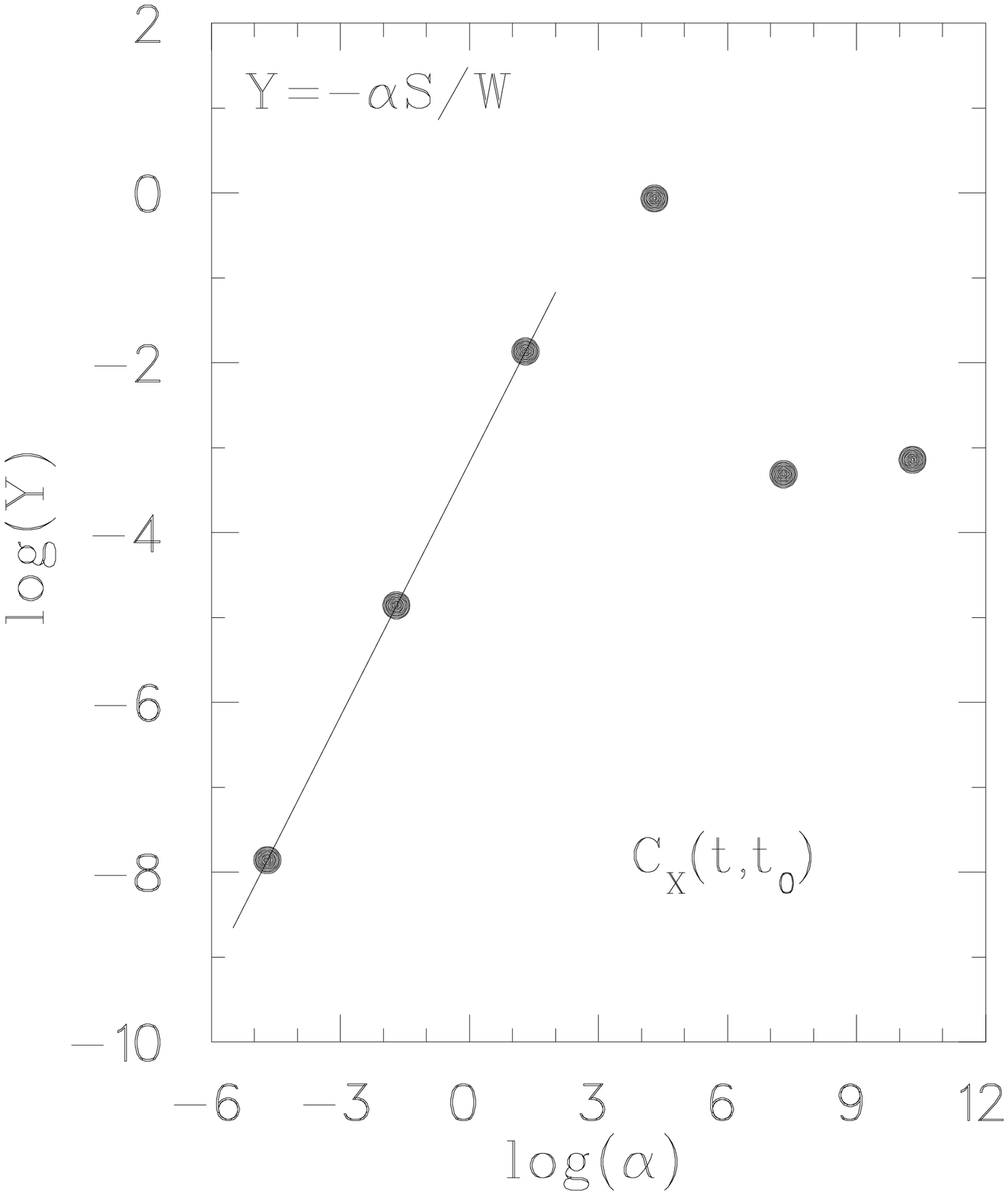}
\includegraphics[angle=0,width=42mm]{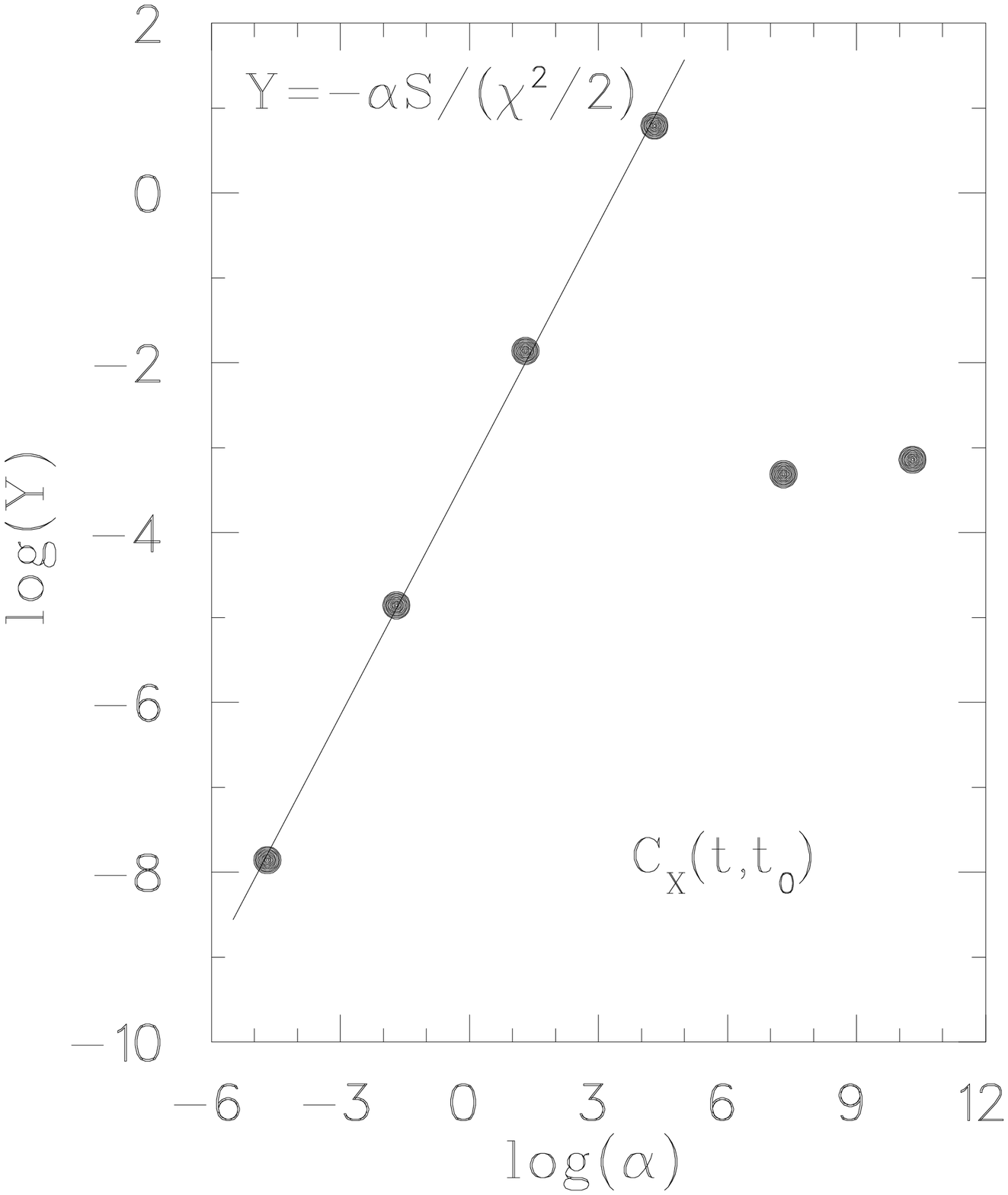}
\caption{\label{fig2}Empirical dependence of the entropy loads $Y_{S/W}$ and
$Y_{S/\chi^2}$ on the entropy weight parameter $\alpha$,
see (\protect\ref{SW}, \protect\ref{SC}).
These results are for the mock
correlator $C_X(t,t_0)$. The lines indicate the extent of linear relationships.}
\end{figure}

In order to decide on a tuning criterion for $\alpha$ it is useful to monitor
quantities like
\begin{eqnarray}
Y_{S/W}&=&\frac{\langle -\alpha S\rangle_{\beta_W\rightarrow\infty}}
{\langle W\rangle_{\beta_W\rightarrow\infty}}\label{SW}\\
Y_{S/\chi^2}&=&\frac{\langle -\alpha S\rangle_{\beta_W\rightarrow\infty}}
{\langle \chi^2/2\rangle_{\beta_W\rightarrow\infty}}\label{SC}\,,
\end{eqnarray}
where $\langle\ldots\rangle_{\beta_W\rightarrow\infty}$
refers to the annealing average measured at the final cooling temperature, $\beta_1$.
We will refer to the above quantities as entropy loads.
Those are shown in Fig.~\ref{fig2}.
It turns out that $\log(Y)$ depends linearly on $\log(\alpha)$
in the regions $\log(\alpha)<+1$ and $\log(\alpha)<+4$,
for $Y_{S/W}$ and $Y_{S/\chi^2}$, respectively.
(In fact $Y\approx 6.2\times10^{-4}\alpha$.) 
Beyond the linear region too much entropy is loaded into the annealing action $W$,
leading to a smoothing of peaks, as seen in Fig.~\ref{fig1}.
Empirically, the criterion emerging from this observation is to tune the entropy weight
such that $\log(Y)\approx -2\pm 1$ within the linear region.
The precise value of $\log(Y)$ is not important, also $Y=Y_{S/W}$ and
$Y=Y_{S/\chi^2}$ work equally well.
As is evident from Fig.~\ref{fig1} results are extremely robust against
varying $\alpha$.

\subsection{\label{sec:single}Single-meson spectrum}

The correlation function
$c(t,t_0)=\langle\hat{\phi}^\dagger(t)\hat{\phi}(t_0)\rangle$
of a single pseudoscalar heavy-light meson operator
$\phi(t)=\sum_{\vec{x}}\overline{Q}_A(\vec{x}t)\gamma_5 q_A(\vec{x}t)$
delivers high quality data in this simulation.
We use these to compare with plateau methods and make some observations
relevant to the present stochastic approach to the MEM.

\begin{figure}
\includegraphics[angle=90,width=84mm]{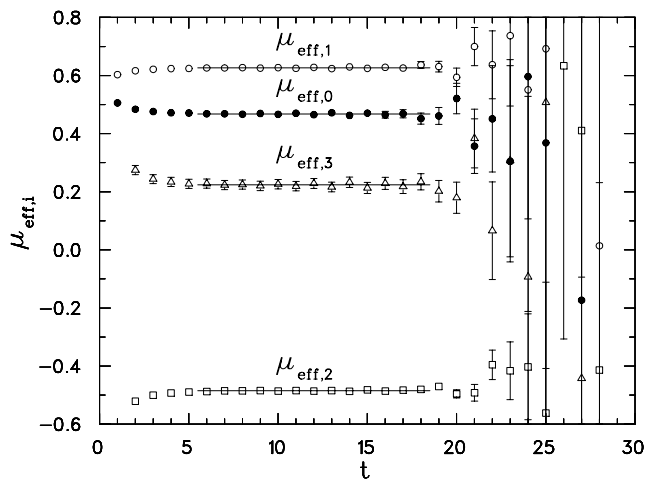}
\caption{\label{fig3}Effective mass functions
(\protect\ref{eff0}--\protect\ref{eff3}) for a single heavy-light meson.
The horizontal lines are plateau fits in the time slice range $6\leq t\leq 18$.}
\end{figure}

In Fig.~\ref{fig3} plots of the mass function discretizations
(\ref{eff0}--\ref{eff3}), built from $c(t,t_0)$, and the corresponding
plateau fits are displayed.
Plateau fits were made directly to $\mu_{\rm eff,i=0\ldots 3}$. 
The resulting masses, other than $m_{\rm eff,0}$, are from solving
(\ref{eff1}--\ref{eff3}). Table~\ref{tab1} shows that those are
consistent within statistical (jackknife) errors.
\begin{table}
\caption{\label{tab1}Plateau masses derived from (\protect\ref{eff0}--\protect\ref{eff3})
on the time slice range $6\leq t\leq 18$.
The entry $E_1$ is the Bayesian result with $\Delta_1$ being the peak width
(standard deviation) computed from the spectral density function $\rho$.
Statistical errors are derived from a gauge configuration jackknife analysis.}
\begin{ruledtabular}
\begin{tabular}{cccccc}
 $m_{\rm eff,0}$ & $m_{\rm eff,1}$ & $m_{\rm eff,2}$ & $m_{\rm eff,3}$ & $E_1$ & $\Delta_1$ \\
\colrule
 0.468(8) & 0.468(7) & 0.468(3) & 0.47(2) & 0.471(15) & 0.017(6)
\end{tabular}
\end{ruledtabular}
\end{table}

\begin{figure}
\includegraphics[angle=0,width=56mm]{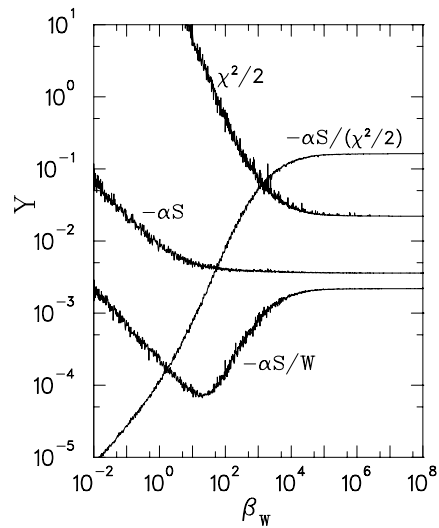}
\caption{\label{fig4}Annealing dynamics in terms of the tuning functions
$Y_{S/W}$, $Y_{S/\chi^2}$, and $Y_{S}$, $Y_{\chi^2}$,
versus the cooling parameter $\beta_W$.
The graphs are labeled with reference to the entropy loads (\protect\ref{SW}, \ref{SC}),
and (\protect\ref{YS}, \ref{YC}).
This example is for the single-meson correlator, with entropy strength
$\alpha=5.0\times 10^{-5}$ and a constant default model $m=1.0\times 10^{-12}$.}
\end{figure}

Figure~\ref{fig4} gives a sense of the annealing dynamics. Beside
(\ref{SW}) and (\ref{SC}) also shown are
\begin{eqnarray}
Y_{S}&=&\langle -\alpha S\rangle_{\beta_W\rightarrow\infty}\label{YS}\\
Y_{\chi^2}&=&\langle \chi^2/2\rangle_{\beta_W\rightarrow\infty}\,.\label{YC}
\end{eqnarray}
In \cite{Fiebig:2001nn} and \cite{Fiebig:2001mr} the use of $Y_{S/W}$ was advocated as a
tuning criterion. In view of Fig.~\ref{fig4} $Y_{S/\chi^2}$ appears to be a better
choice given its monotonic nature.
A target entropy load of $Y_{S/\chi^2}\approx 10^{-1\pm 1}$ is a safe tuning criterion,
provided the cooling algorithm runs in the (upper) linear region, see Fig.~\ref{fig2}.
 
\begin{figure}
\includegraphics[angle=0,width=42mm]{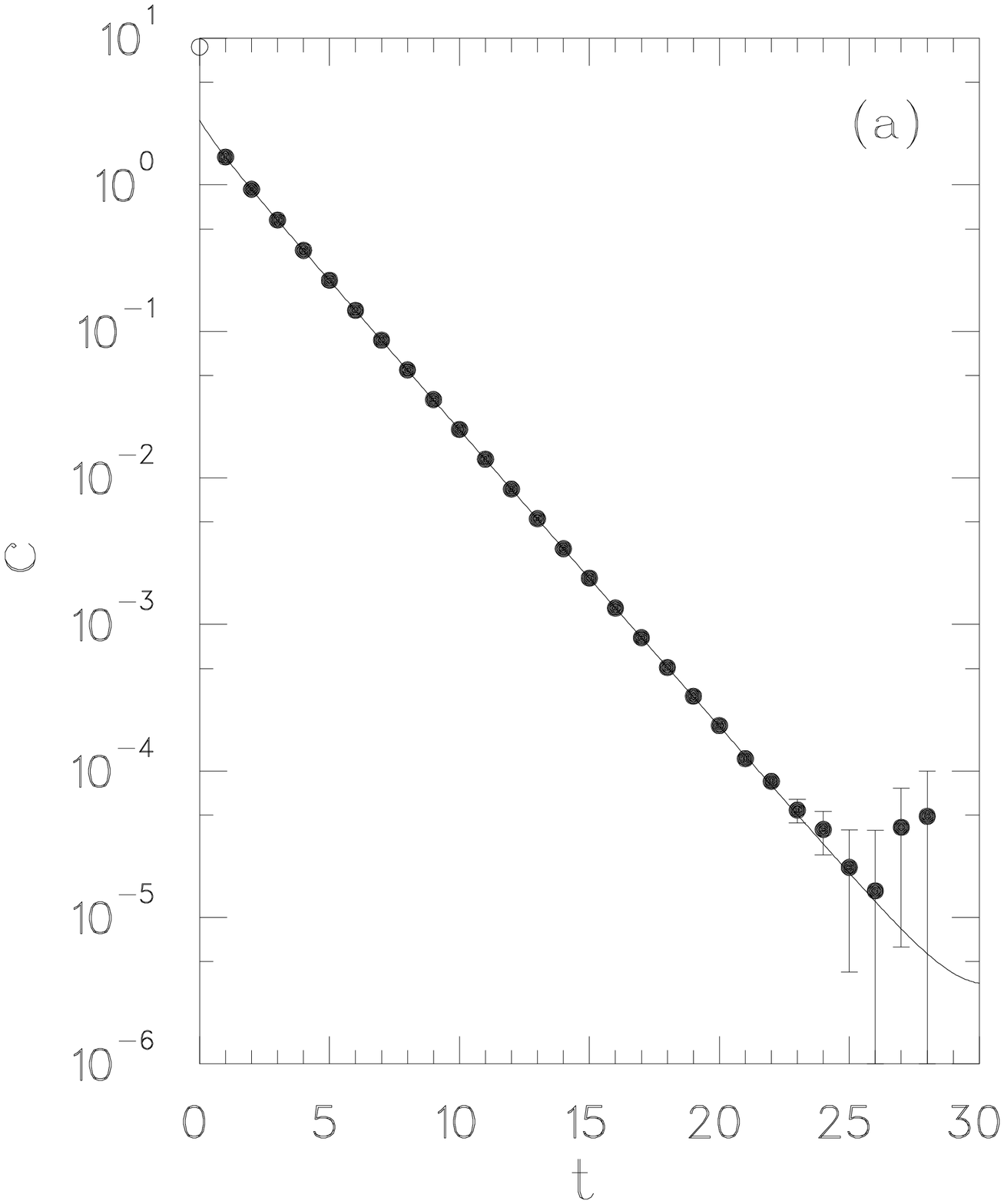}
\includegraphics[angle=0,width=42mm]{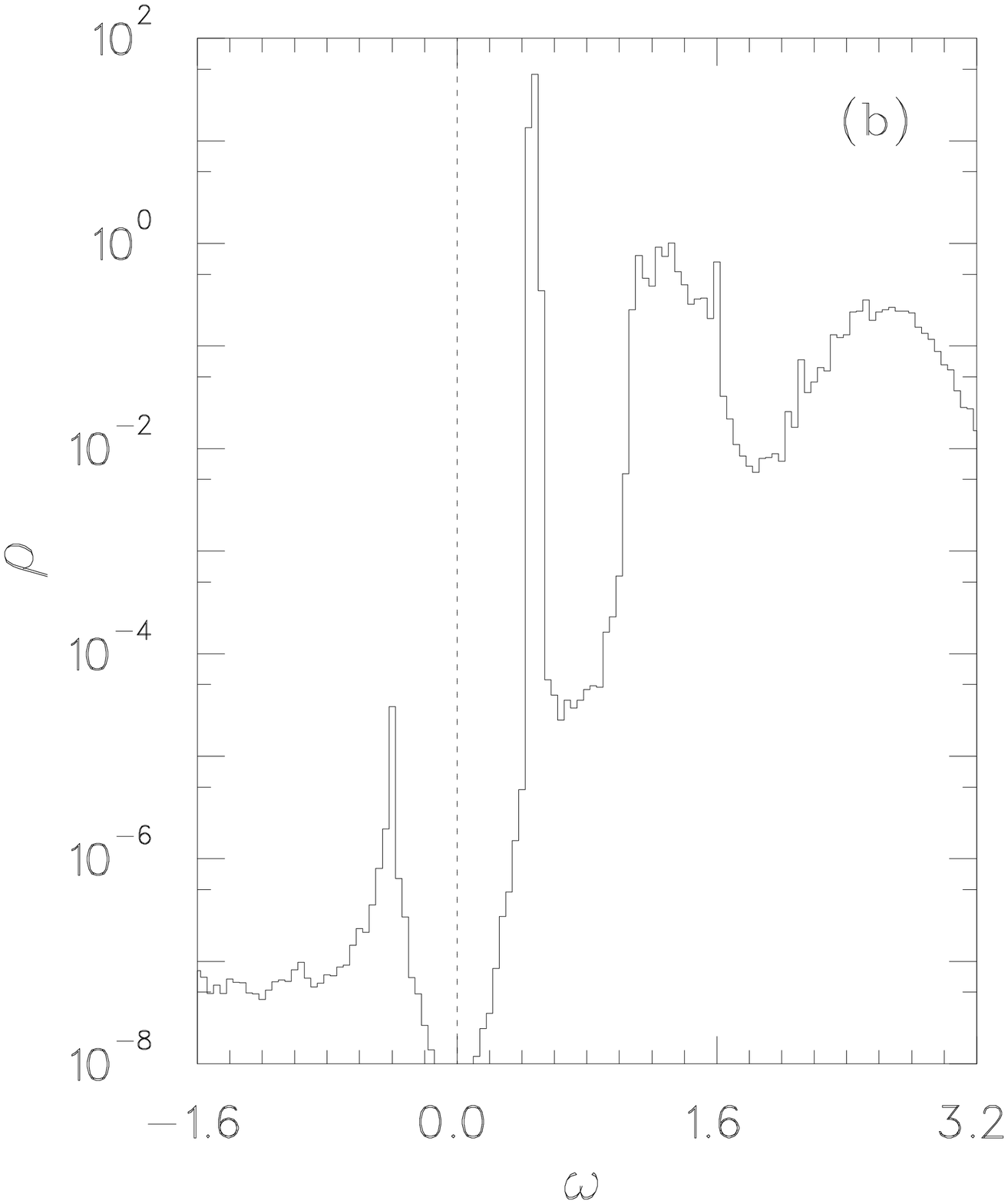}
\caption{\label{fig5ab}Time correlation function for a single heavy-light meson
together with a Bayesian fit (a),
and the corresponding spectral density function (b).
This result stems from a single random start, with entropy weight
$\alpha=5.0\times 10^{-5}$,
and a constant default model $m=1.0\times 10^{-12}$.} 
\end{figure}

The Bayesian analysis of the time correlation function $c(t,t_0)$ is
shown in Fig.~\ref{fig5ab}.
The solid line in Fig.~\ref{fig5ab}(a) derives from the computed spectral
density $\rho$, via (\ref{Fd}).
With the exception of $t_0=0$ all available time slices were used.
Parameters are $\alpha=5.0\times 10^{-5}$, for the entropy strength,
a constant default model $m=1.0\times 10^{-12}$,
and a random annealing start about $m$.
The graph of $\rho$ in Fig.~\ref{fig5ab}(b) exhibits a global structure
consisting of distinct peaks, some broad, and a micro structure of fluctuations
on the scale of $\Delta\omega$.
The micro structure depends on details of the annealing
process, particularly the start configuration.
Clearly, it makes no sense to infer the micro structure from the data.
The reason is that only $T-1=29$ data points do not contain
enough information to determine $K_{+}-K_{-}+1=K=121$
spectral parameters (with any sizable probability).

On the other hand the global structure is a stable feature.
In the region $\omega>0$ three peaks can be distinguished
in Fig.~\ref{fig5ab}(b). By way of inspection we loosely define
\begin{equation}
\delta_n=\{\omega:\omega\in{\rm peak}\ \#n\}\quad n=1,2\ldots\,.
\label{deltan}\end{equation}
Then, for each peak $n$, we may calculate the volume $Z_n$,
the mass $E_n$, and the width $\Delta_n$, according to
\begin{eqnarray}
Z_n&=&\int_{\delta_n}d\omega\/\rho_T(\omega)\label{Zn}\\
E_n&=&Z_n^{-1}\int_{\delta_n}d\omega\/\rho_T(\omega)\omega\label{En}\\
\Delta_n^2&=&Z_n^{-1}\int_{\delta_n}d\omega\/\rho_T(\omega)\left(\omega-E_n\right)^2\,.
\label{Dn}\end{eqnarray}
These integrated, low moment, quantities are evidently insensitive to
the micro structure. They constitute the information that reasonably can
be expected to flow from the Bayesian analysis.

\begin{figure}
\includegraphics[angle=0,width=41.2mm]{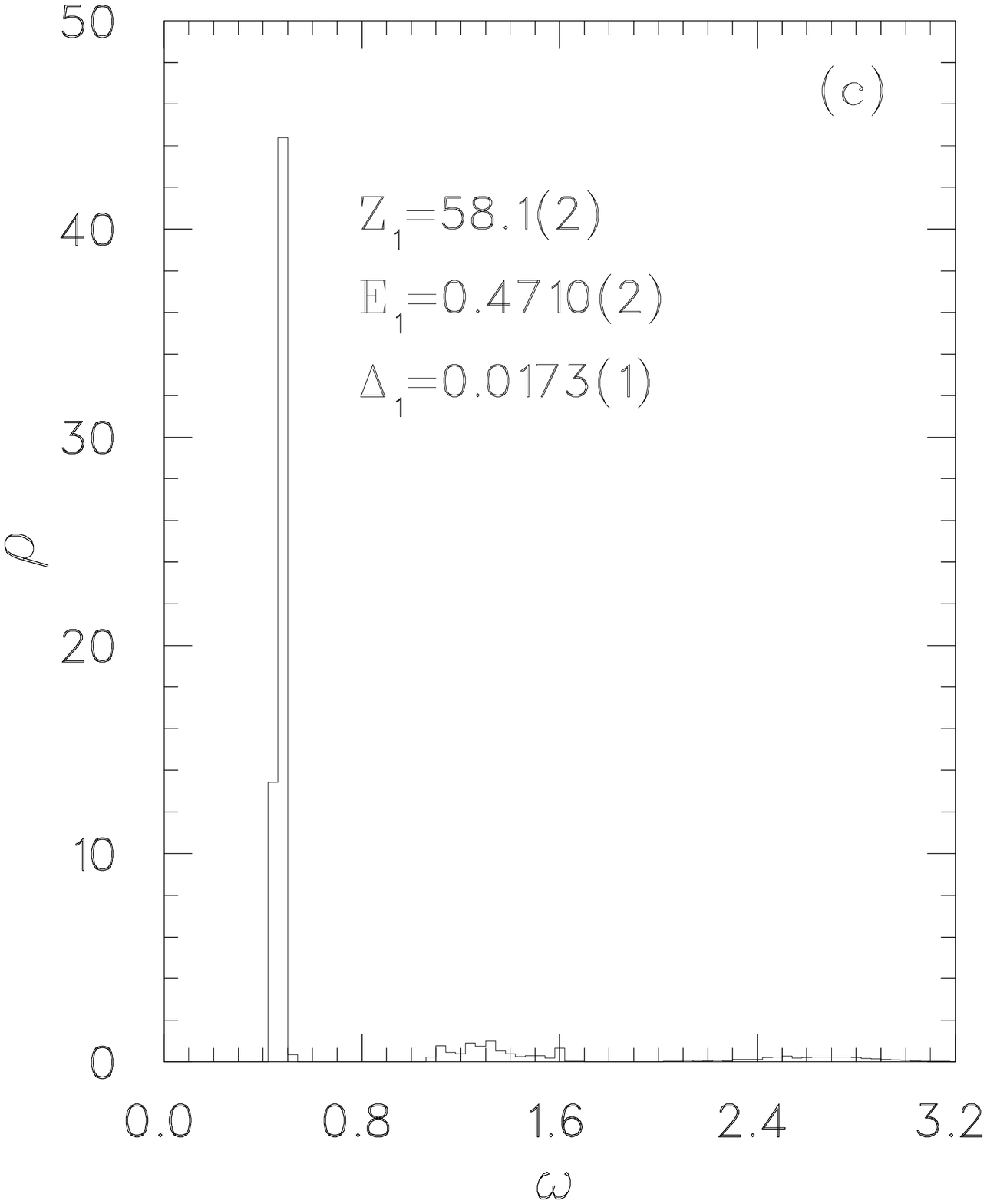}
\includegraphics[angle=0,width=42.8mm]{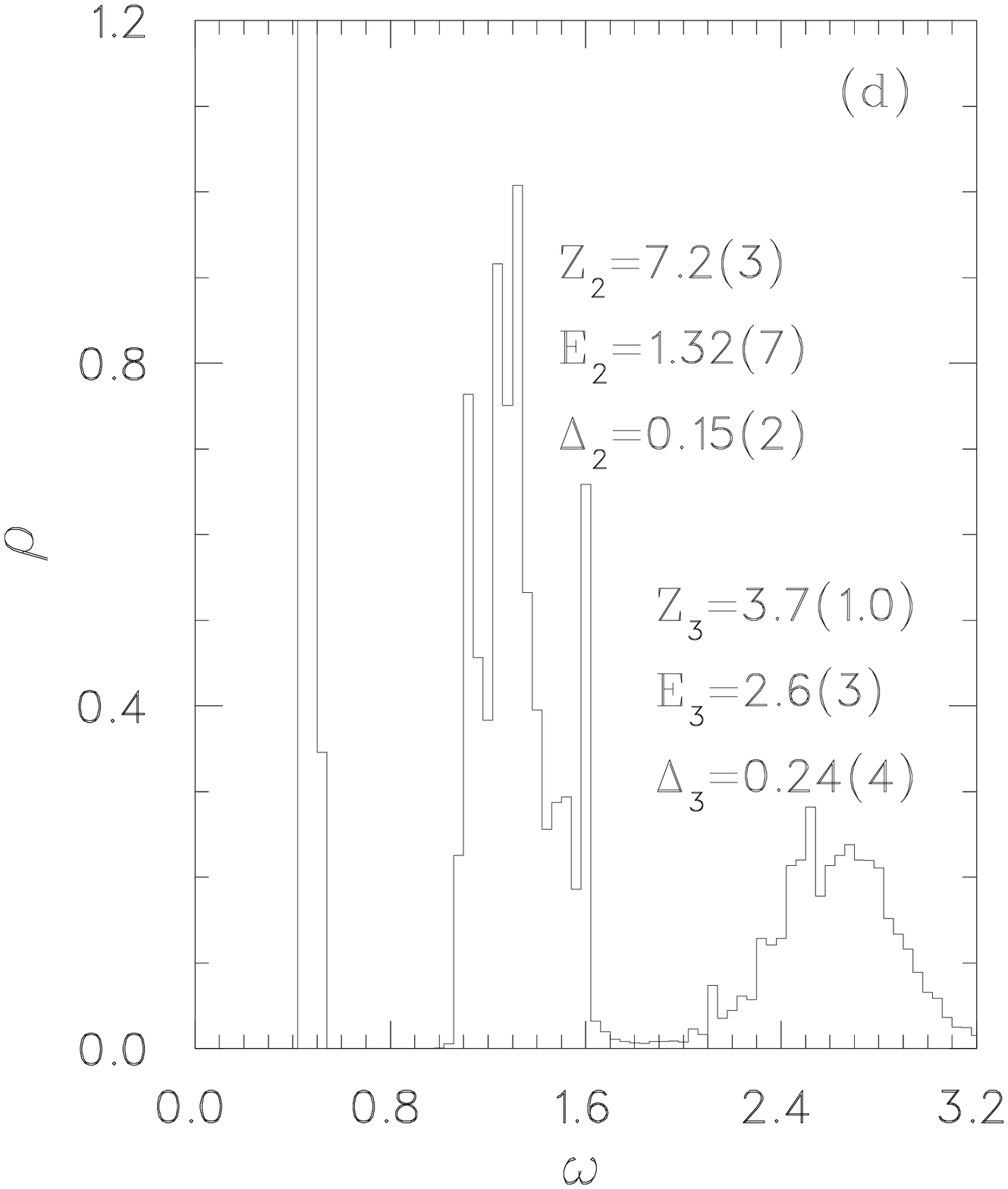}
\caption{\label{fig5cd}Spectral density $\rho$ for a single heavy-light meson,
same as in Fig.~\protect\ref{fig5ab}(b), but on linear scales, emphasizing
the ground and the excites states (c) and (d), respectively.
The uncertainties of $Z_n$, $E_n$, and $\Delta_n$ are standard deviations
from eight annealing runs.}
\end{figure}

The spectral density of Fig.~\ref{fig5ab}(b) is replotted in Fig.~\ref{fig5cd} on
linear scales. The tall narrow peak in Fig.~\ref{fig5cd}(c) corresponds to the
plateau masses of Fig.~\ref{fig3}, as listed in Tab.~\ref{tab1}.
There, the entries $E_1$ and $\Delta_1$ are the Bayesian results.
Their statistical errors are derived from
a jackknife analysis selecting four subsets of gauge configurations.
(Note that the uncertainties in Fig.~\ref{fig5cd} are standard
deviations from eight annealing starts.)
Cold starts from the default model $m$ were used
to suppress the dependence on the annealing start configuration. 
The peak width $\Delta_1$ is comparable to the gauge configuration statistical error.
This is the exception. With correlation function data of lesser quality
(like with the two-meson operators below)
the size of the peak width is typically larger than the statistical error.
It appears that the peak width $\Delta_n$ is related to the size $\Theta_n$
of the corresponding effective mass function plateau, like in Fig.~\ref{fig3},
or the size of the $\log$-linear stretch in a plot like in Fig.~\ref{fig5ab}(a).
As a very coarse description $\Delta_n \Theta_n\approx {\rm const}$
comes to mind. Using $\Theta_1=12$ and $\Delta_1=0.017$ we have
${\rm const}\approx 0.2$.
The peaks $n=2$ and $n=3$ seen in Fig.~\ref{fig5cd}(d) would thus appear to
originate from $\Theta_n\approx 0.2/\Delta_n$, or 1.3 and 0.8 time slices,
respectively.
(By inspection of Fig.~\ref{fig3} as many as 5 time slices appear involved,
however.)
The physical relevance of, at least, peak $n=3$ is therefore questionable.
On the other hand it is remarkable that the maximum entropy method is
sensitive to the slightest details in the correlation function data.

\subsection{\label{sec:defaultm}Default model dependence}

The Shannon-Jaynes entropy (\ref{Smem}) implies the possible dependence of
the computed spectral density $\rho$ on the default model
$m=\{m_k : K_- \leq k \leq K_+\}$.
We explore the $m$ dependence using as an example the time correlation function
$C_v$ with $v_1=1$ $v_2=0$, in the notation of (\ref{Phiv}), at relative distance $r=4$.

\begin{figure}
\includegraphics[angle=0,width=42.2mm]{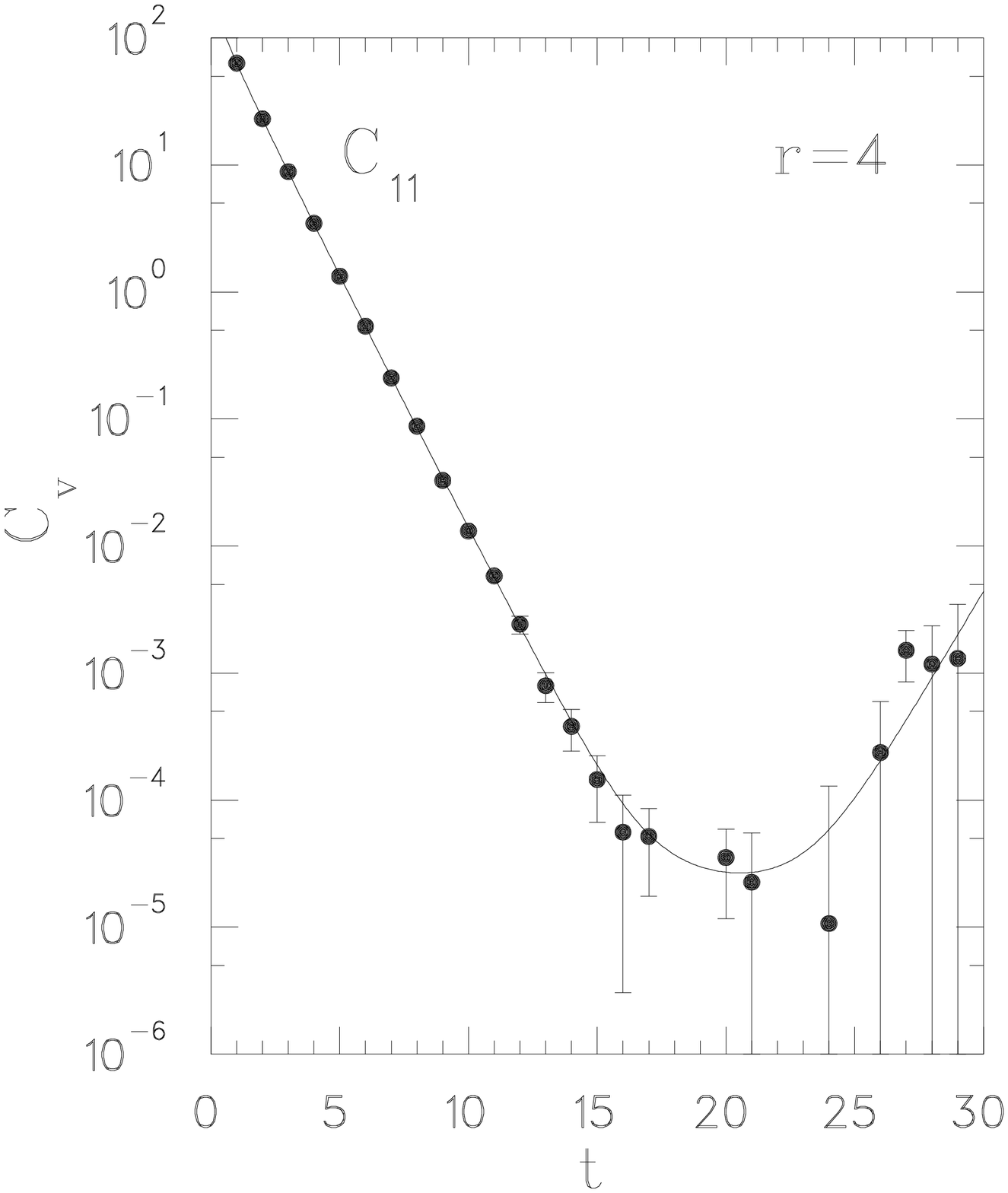}
\includegraphics[angle=0,width=41.8mm]{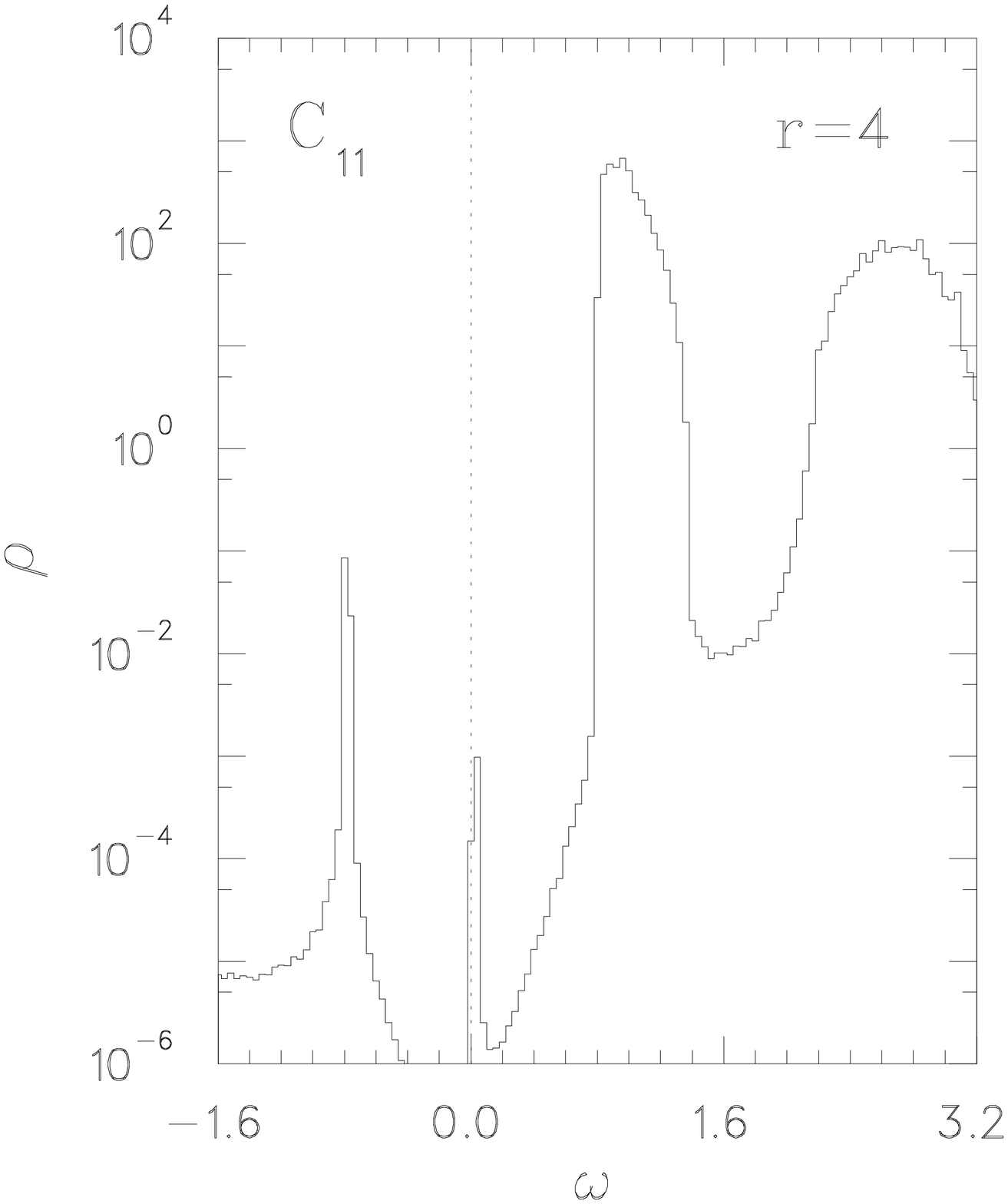}
\caption{\label{fig6d7d}Correlation function
$C_{11}=\langle\hat{\Phi}^\dagger_1(t)\hat{\Phi}_1(t_0)\rangle$
of a heavy-light meson-meson operator
at relative distance $r=4$. The Bayesian fit (solid line) is from the spectral
density $\rho$ shown on the right. At $\alpha=2\times 10^{-6}$ and constant default
model $m=1.0\times 10^{-12}$ the spectral density $\rho$ is obtained from an
average over eight random annealing start configurations.
The average entropy load is $Y_{S/\chi^2}=0.477$ for these runs.}
\end{figure}

Figure~\ref{fig6d7d} shows the time correlation function data together with the
Bayesian fit, and the corresponding spectral density $\rho$. The latter
is the average over eight random annealing start configurations.
This has the effect of smoothing out the micro structure of $\rho$.
We have used a constant default model $m_k=1.0\times 10^{-12}$, all $k$.

\begin{figure}
\includegraphics[angle=0,width=42mm]{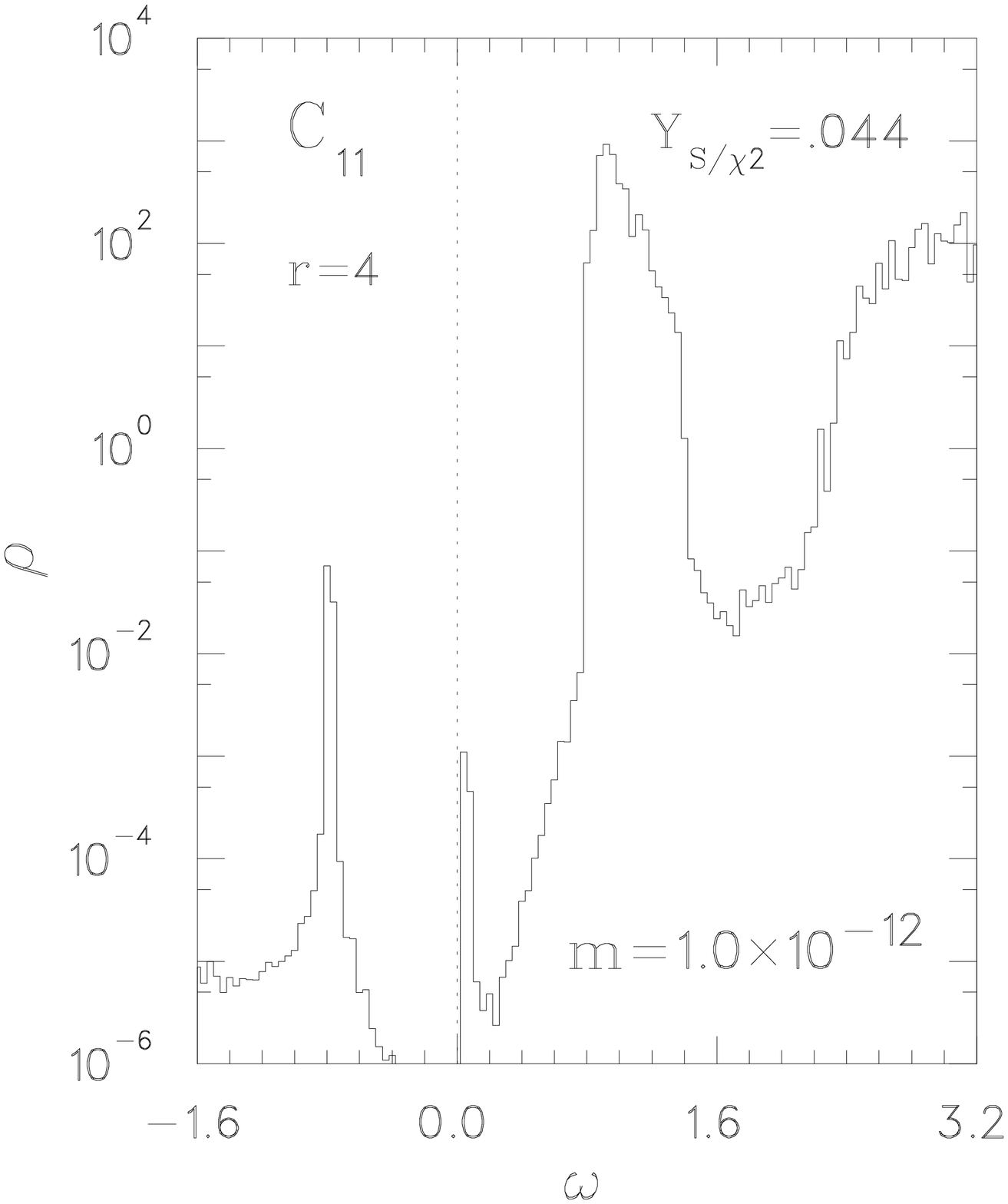}
\includegraphics[angle=0,width=42mm]{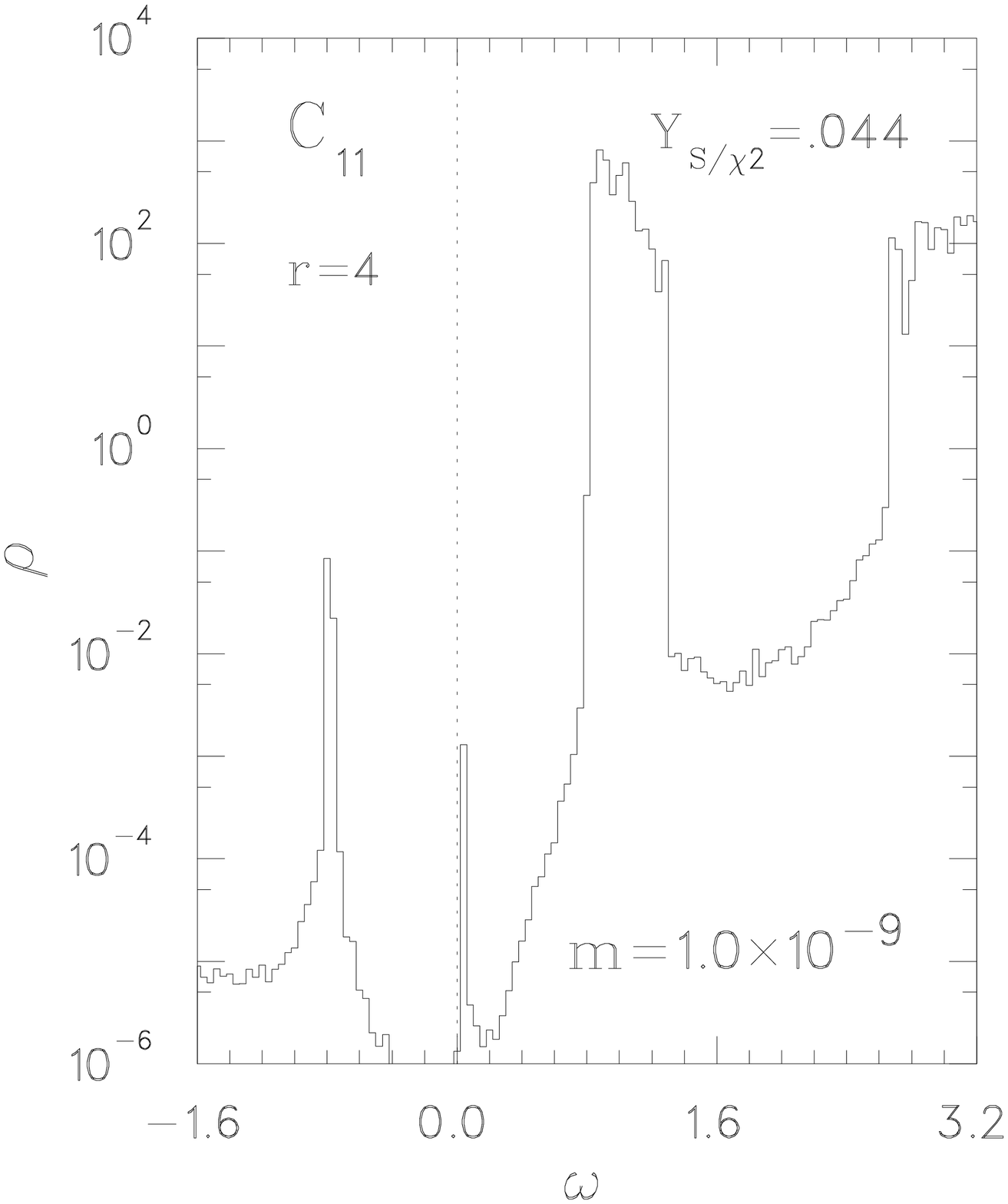} \\
\includegraphics[angle=0,width=42mm]{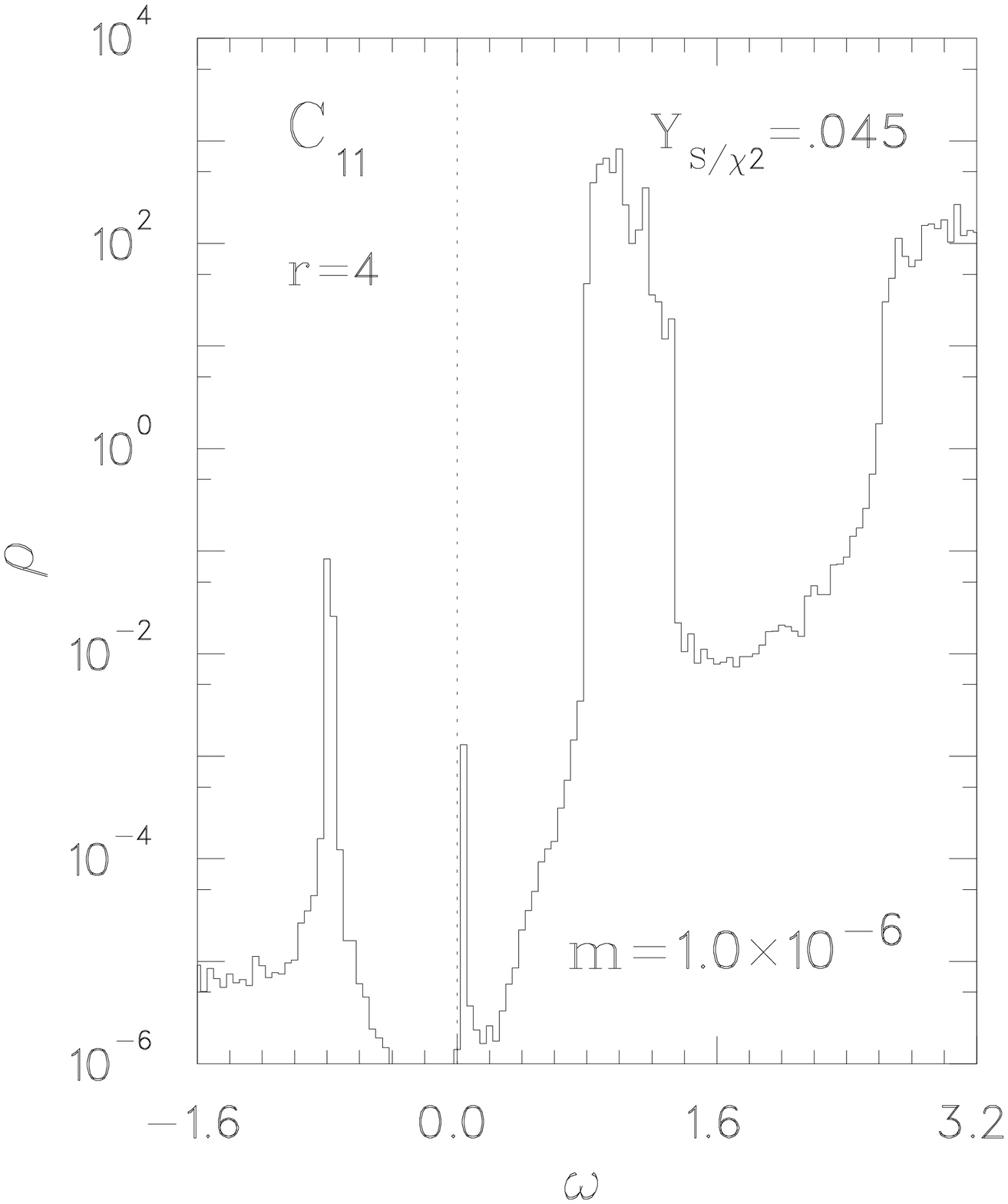}
\includegraphics[angle=0,width=42mm]{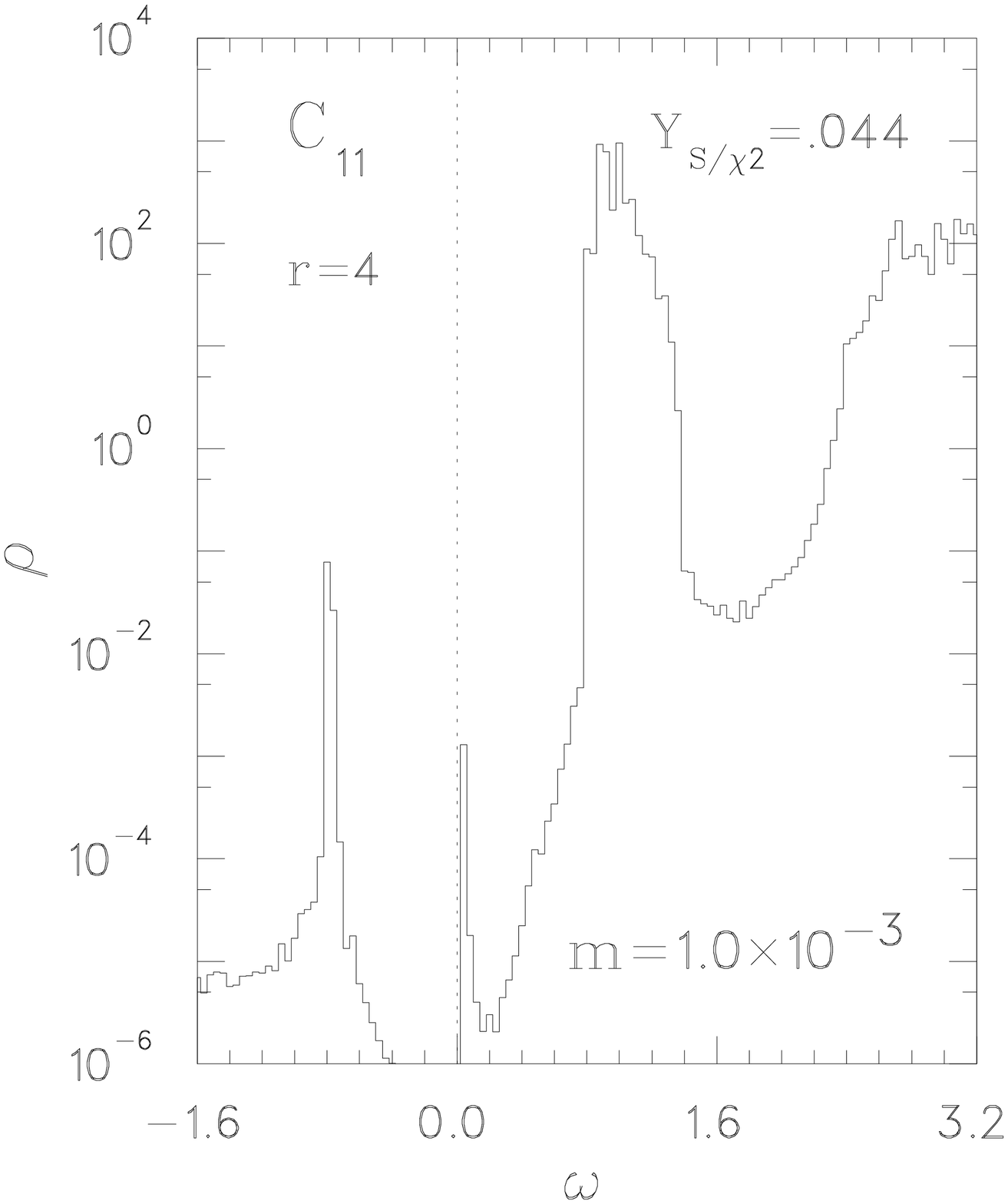} \\
\includegraphics[angle=0,width=42mm]{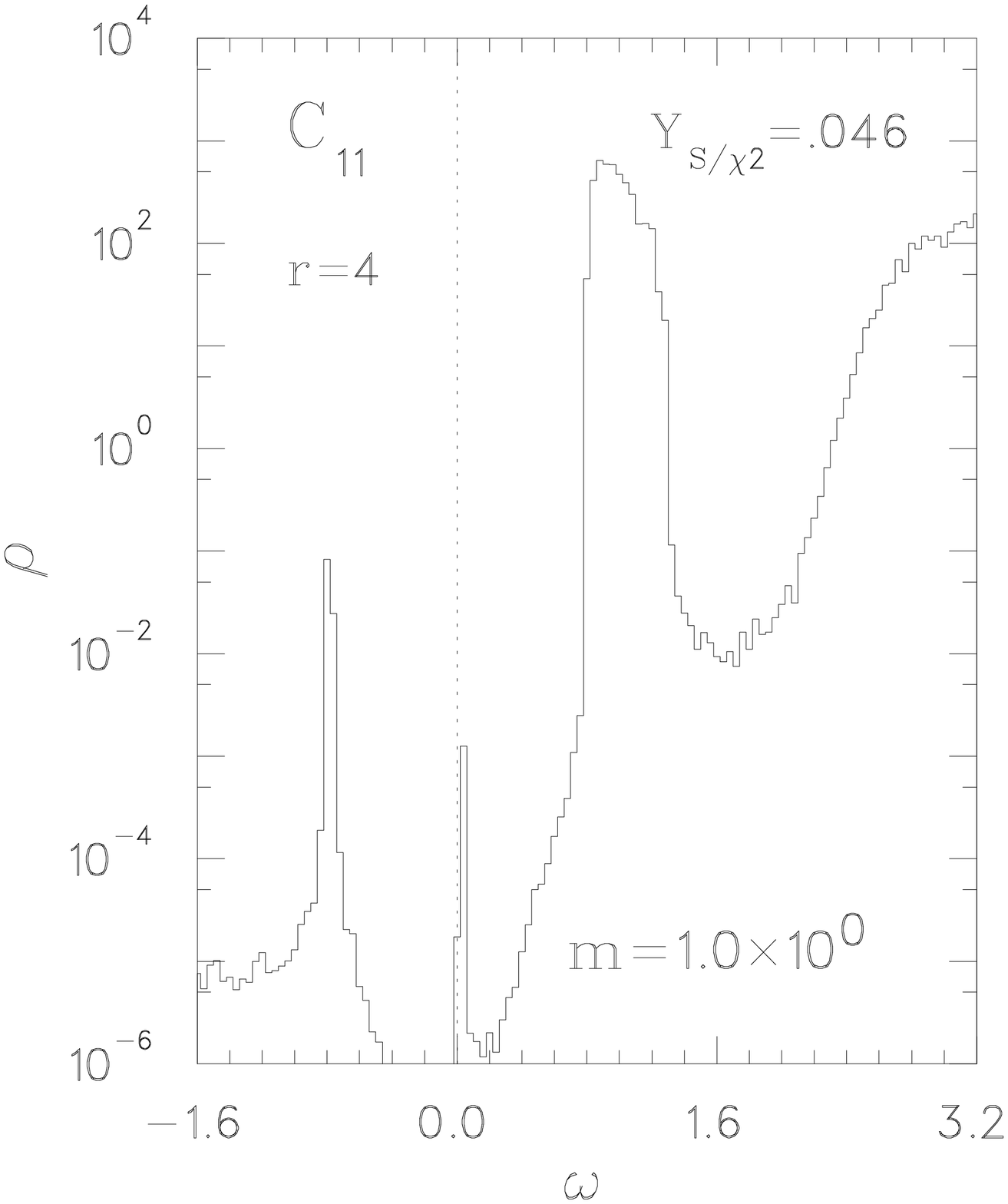}
\includegraphics[angle=0,width=42mm]{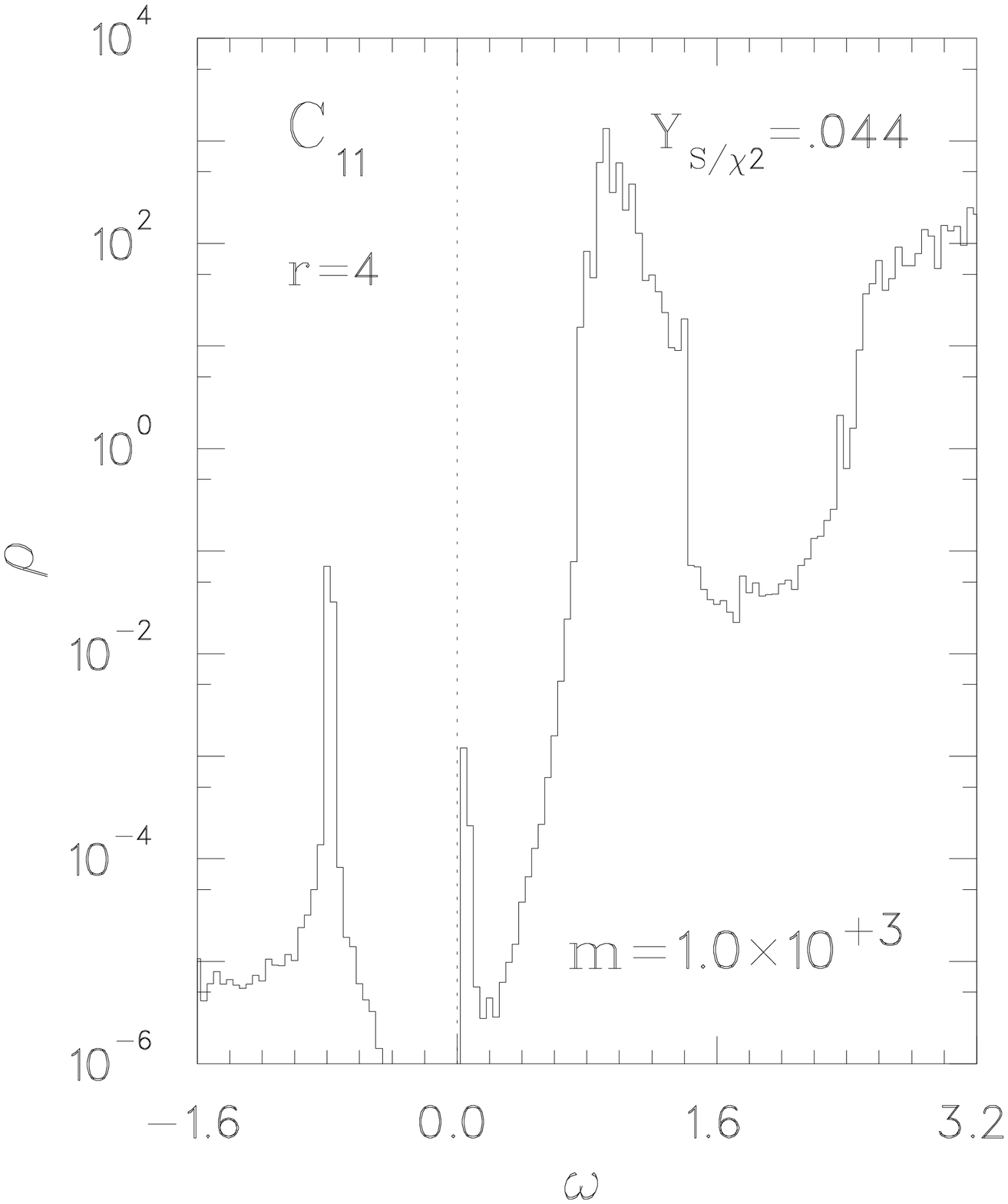}
\caption{\label{fig18}A sequence of spectral densities $\rho$ obtained from a wide range of
constant default models $m$, see inserts. The entropy strength parameter $\alpha$
was tuned to keep the entropy load constant, $Y_{S/\chi^2}\approx 0.045$.
The operator is the same as in Fig.~\protect\ref{fig6d7d}.}
\end{figure}

The stability of this result is tested by varying the default model through
15 orders of magnitude, $m=10^{-12}\ldots 10^{+3}$, as shown in Fig.~\ref{fig18}.
To keep effects of the annealing start configuration small
cold starts from $\rho=m$, using the same random seed, were employed for all values
of $m$. In each case the entropy strength parameter $\alpha$ was tuned such that
the entropy load $Y_{S/\chi^2}$ remained constant. Aside from the familiar
micro structure fluctuations, the global (physical) features are stable
within the range of, a remarkable, fifteen orders of magnitude.
Numerical experiments with non-constant $m$ do not change this assessment.
In Tab.~\ref{tab2} are listed the three integral quantities 
(\ref{Zn})--(\ref{Dn}) averaged over the six default models
together with the corresponding standard deviations. Their smallness (0.3--3\%)
attests to the default model independence of the Bayesian fits.
Given the huge variation of the default model the stability of $\rho$ is remarkable.
\begin{table}
\caption{\label{tab2}Averages of volume, energy, and width of the dominant peak
seen in Fig.~\protect\ref{fig18} over the six default model choices
$m=10^{-12}\ldots 10^{+3}$ at fixed entropy load $Y_{S/\chi^2}\approx 0.045$.
The uncertainties are the corresponding standard deviations.
The entry $m_{\rm eff,0}$ is the plateau mass (\protect\ref{eff0}) from
Fig.\protect\ref{fig3Rab} with the statistical (jackknife) error,
see Sect.~\protect\ref{sec:plateau}.}
\begin{ruledtabular}
\begin{tabular}{cccc}
 $Z_1$ & $E_1$ & $\Delta_1$ & $m_{\rm eff,0}$ \\
\colrule
3923.(18.) & 0.972(3) & 0.100(3) & 0.94(1)
\end{tabular}
\end{ruledtabular}
\end{table}

\subsection{\label{sec:anneal}Annealing start dependence}

The annealing algorithm starts with some initial spectral configuration $\rho_{\rm ini}$.
Depending on the purpose we have used cold starts from the default model, $\rho_{\rm ini}=m$,
or random starts from the default model, $\rho_{{\rm ini},k}=x_km_k$, where the $x_k$ are
drawn from a gamma distribution of order two, $p_a(x)=x^{a-1}e^{-x}/\Gamma(a), a=2$.
The global features of the final spectral density are of course independent of the start
configuration, but the micro structure of $\rho$ is not. The reason is that in practice
the annealing process is neither infinitely slow nor is the final cooling
temperature $\beta_1^{-1}$ exactly zero.
Therefore the annealing result for $\rho$ settles close to the global minimum,
say $\rho_{\rm min}$, of $W[\rho]$.
Considering annealing (thermal) fluctuations only, we expect the deviation
$|\rho-\rho_{\rm min}|$ to be large in directions
(of $\rho$ space) where the minimum is shallow.
Thermal fluctuations are easily controlled, however. Those were kept negligible
in the present study.
More importantly, there may be local minima close to $\rho_{\rm min}$ which
are only slightly larger than $W[\rho_{\rm min}]$.
This situation invites computing a set of spectral densities from different, say random,
initial configurations. The averages and standard deviations of the $\rho_k$
then gives some insight into the structure of the peak and the nature of the minimum of $W$
and its neighborhood. 

\begin{figure}
\includegraphics[angle=0,width=42.2mm]{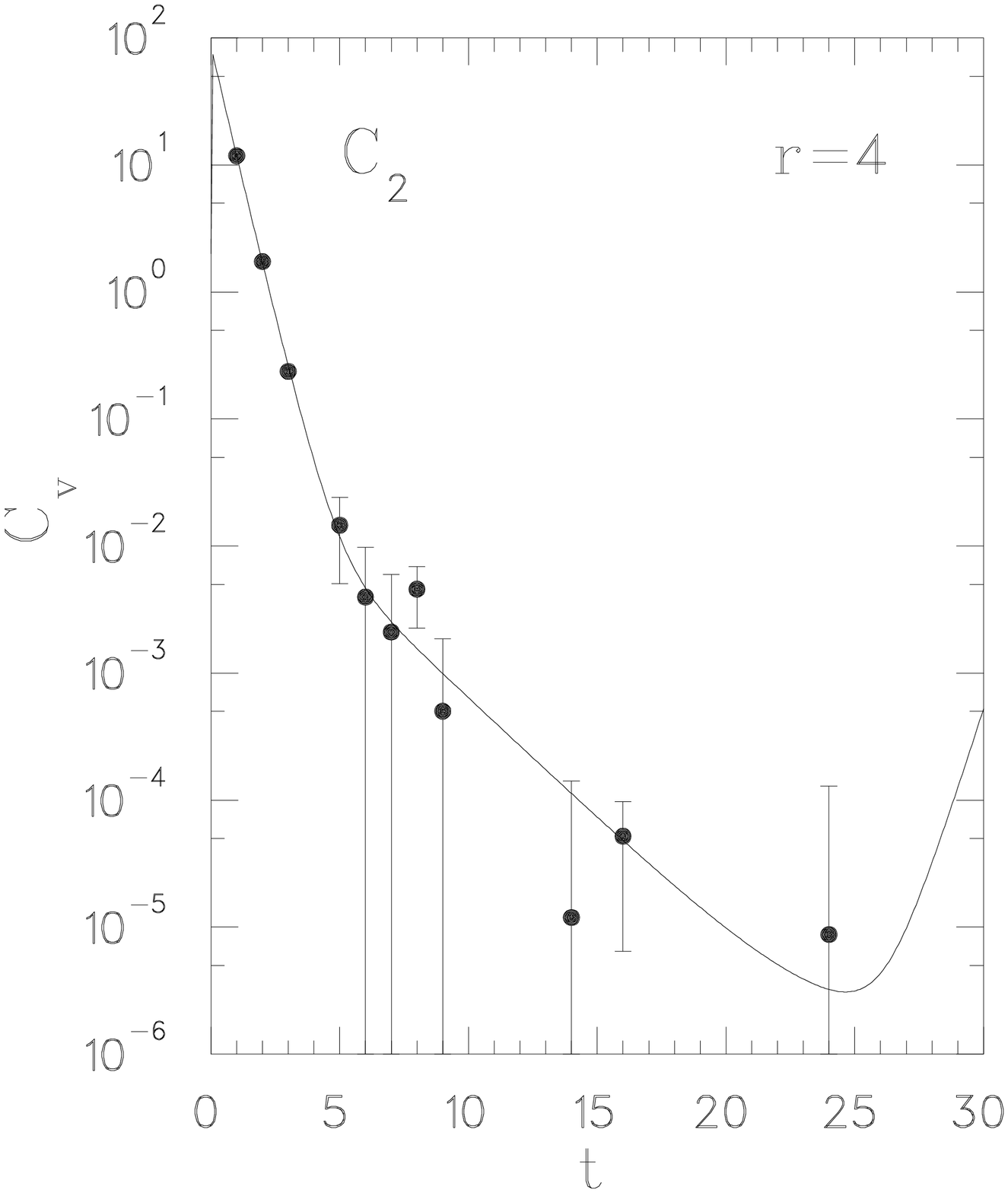}
\includegraphics[angle=0,width=41.8mm]{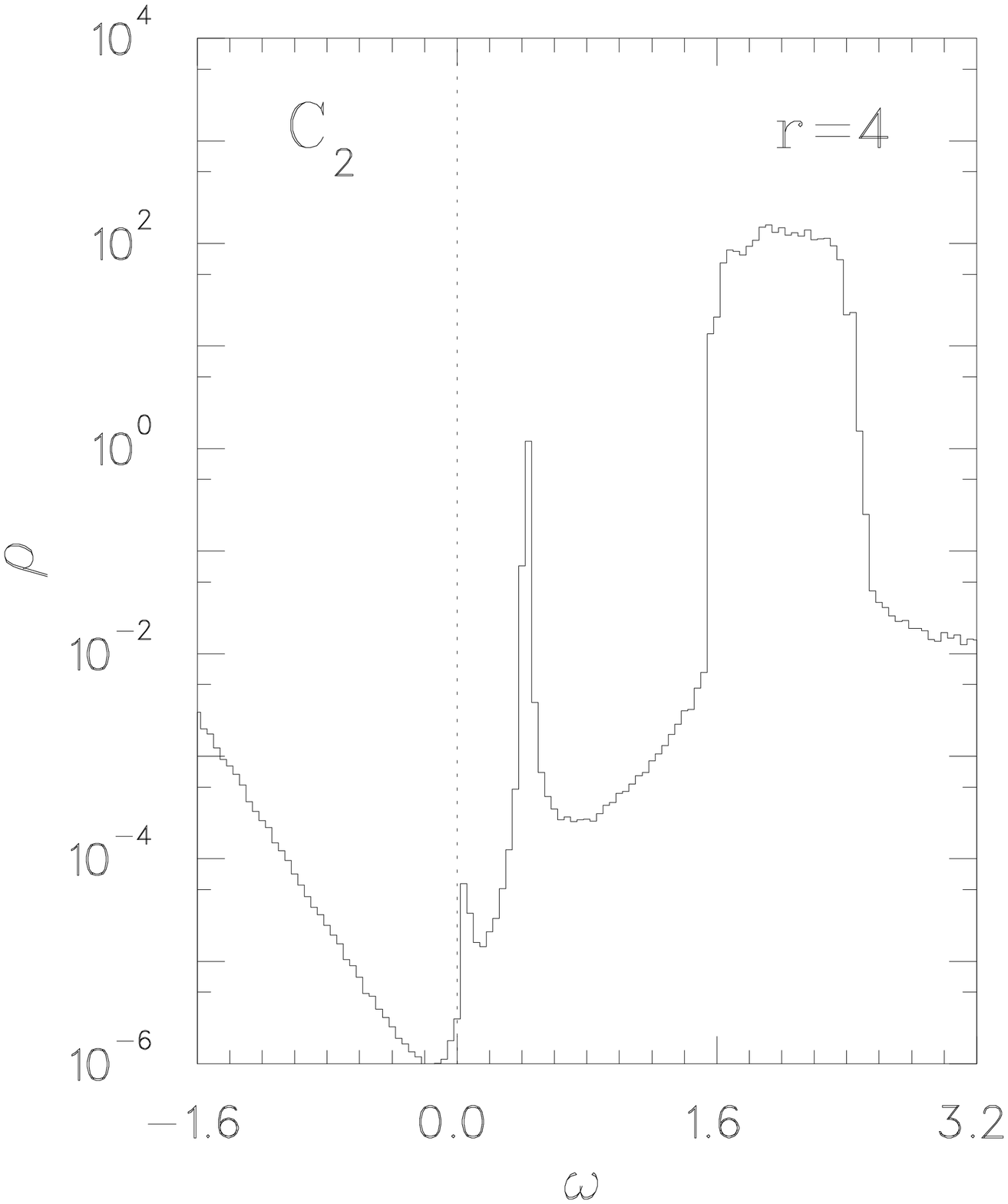}
\caption{\label{fig15d16d}Excited state correlation function $C_2$ of a heavy-light
meson-meson operator
at relative distance $r=4$. The Bayesian fit (solid line) is from the spectral
density $\rho$ shown on the right. At $\alpha=5\times 10^{-7}$ and constant default
model $m=1.0\times 10^{-12}$ the spectral density $\rho$ is obtained from an
average over eight random annealing start configurations.}
\end{figure}

\begin{figure}
\includegraphics[angle=0,width=42mm]{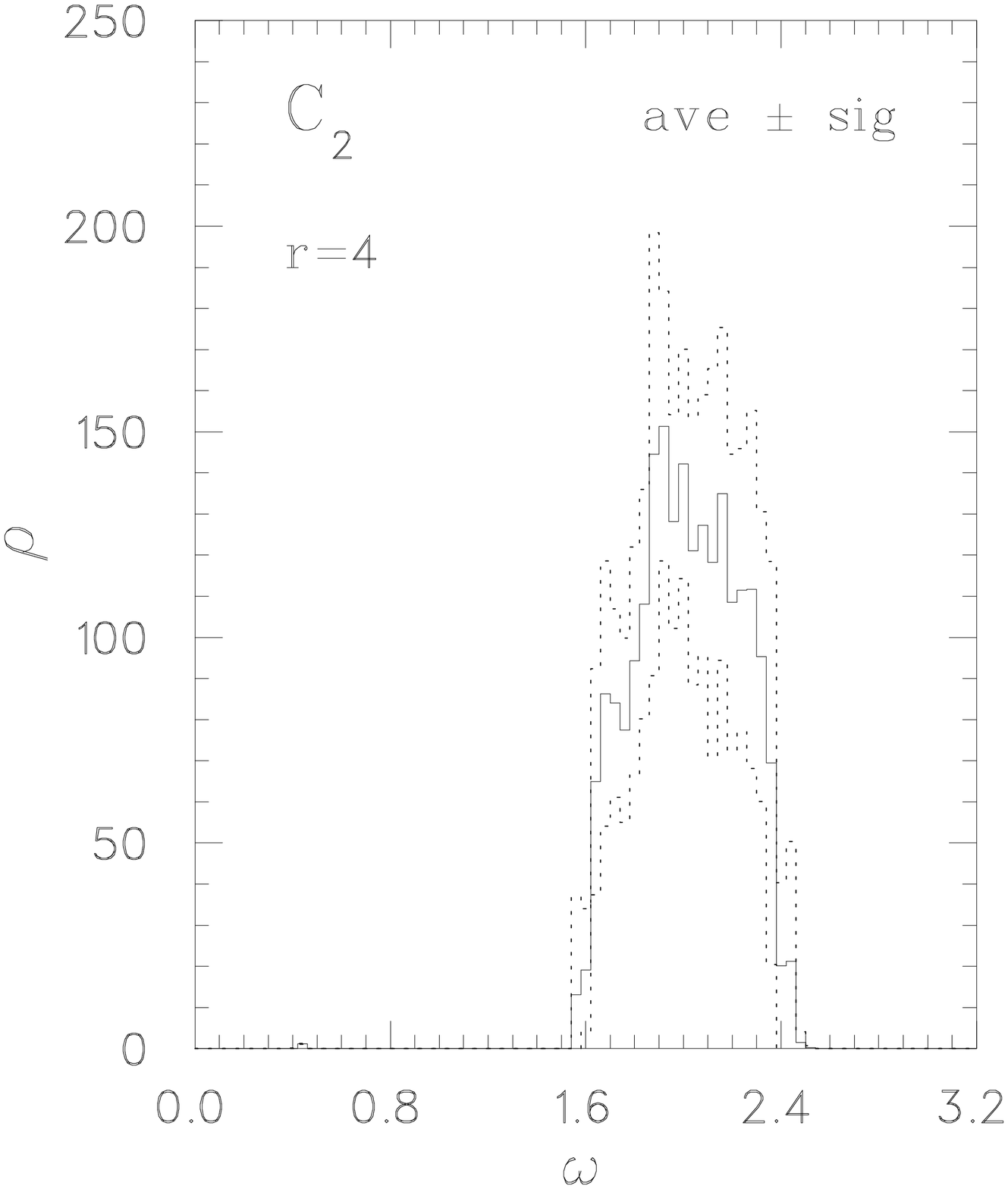}
\includegraphics[angle=0,width=42mm]{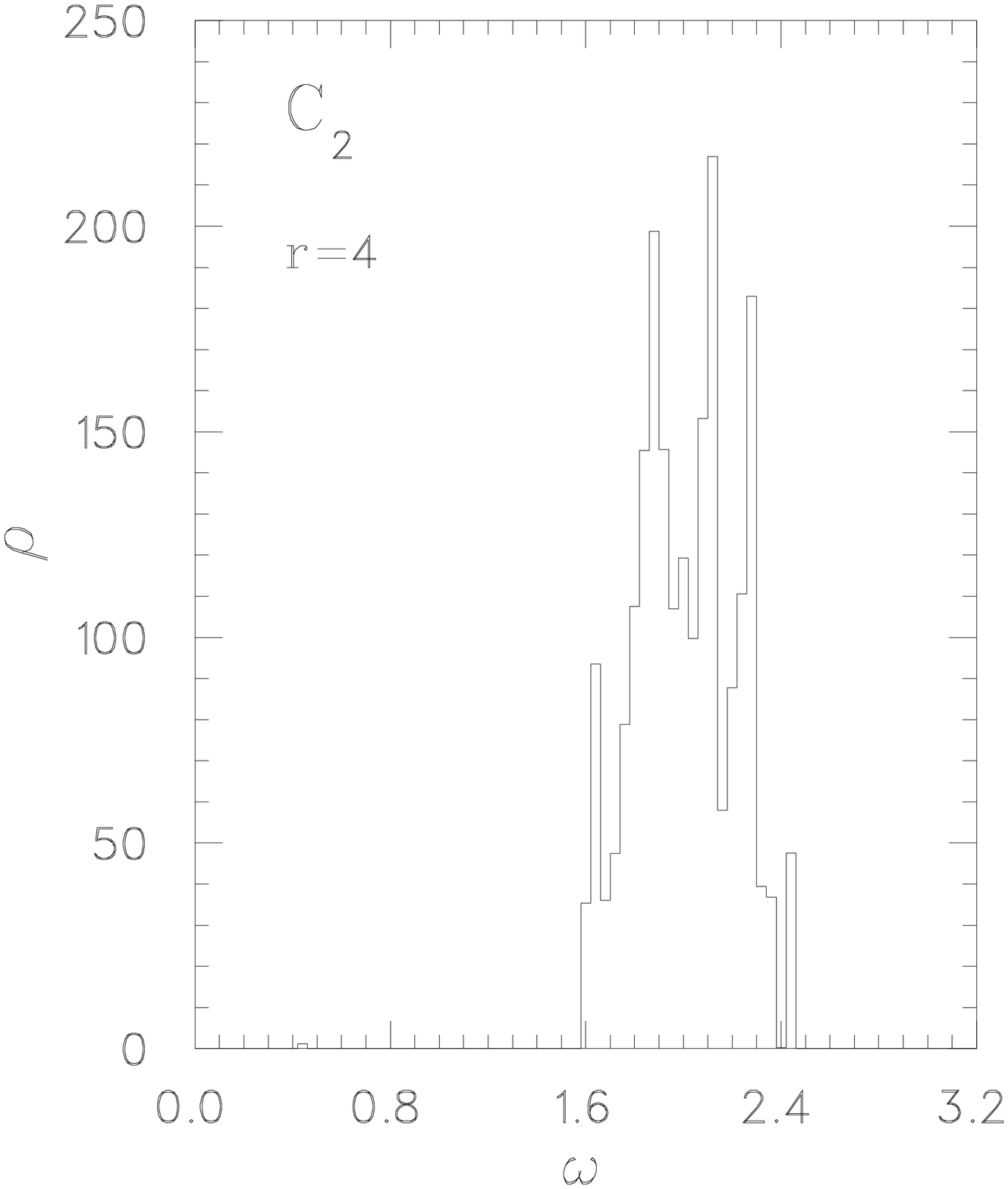} \\
\includegraphics[angle=0,width=42mm]{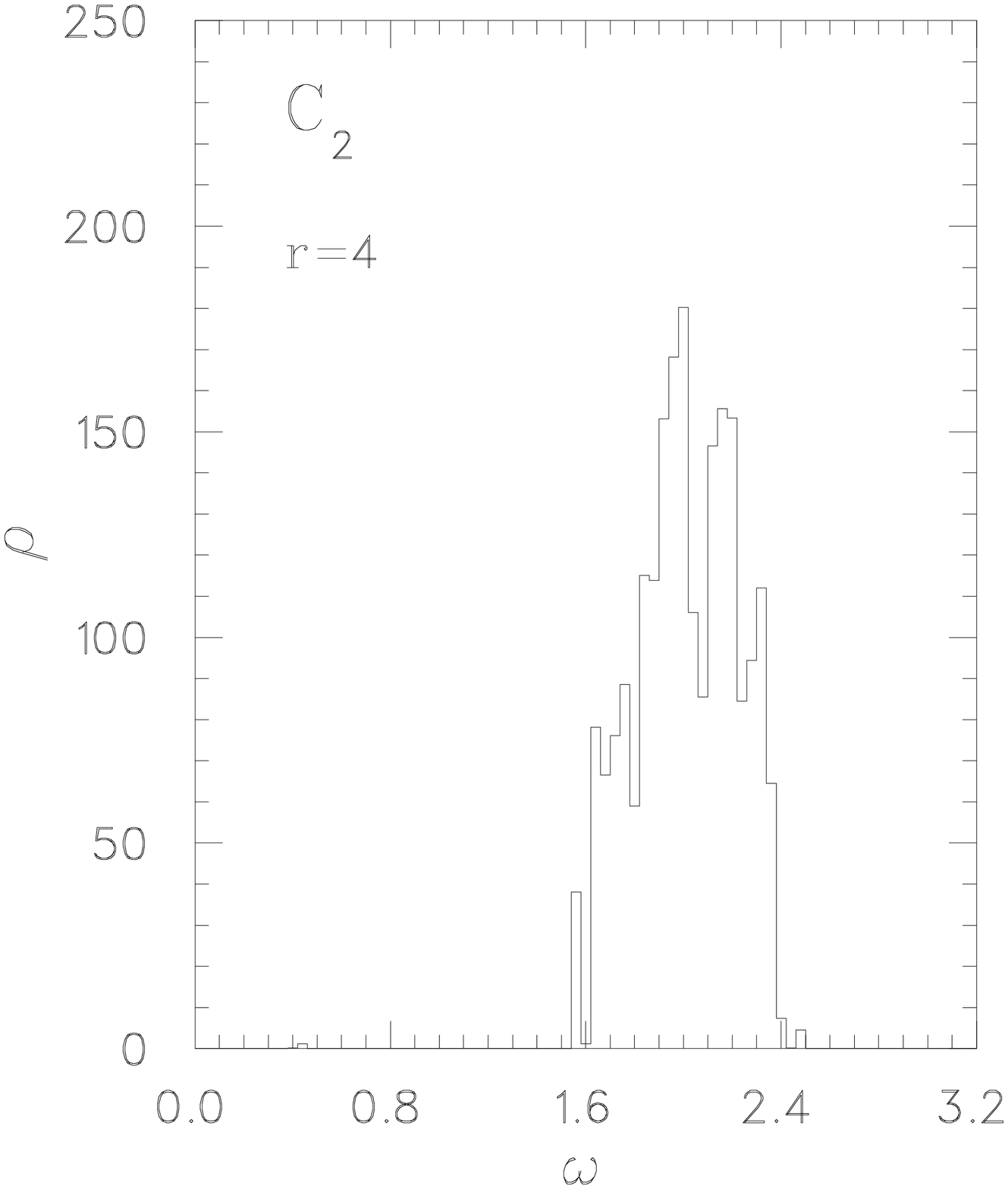}
\includegraphics[angle=0,width=42mm]{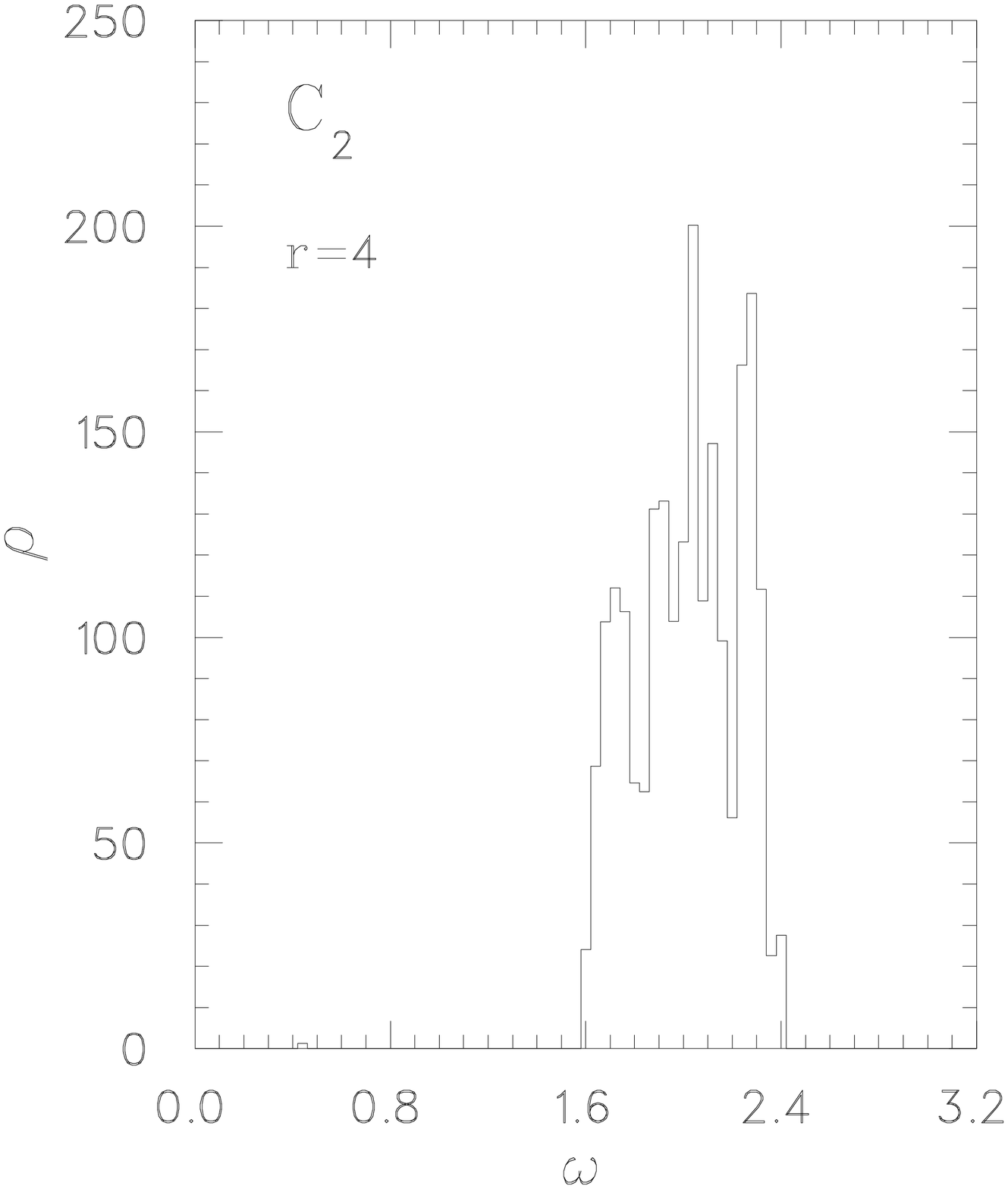}
\caption{\label{fig19}Spectral densities $\rho$ of the excited state correlation
function of Fig.~\protect\ref{fig15d16d}. The sequence of four frames shows the
average (ave) over a sample of eight random annealing start configurations including
the bounds (dotted lines) of one standard deviation ($\pm$sig), and three selected
examples of spectral functions making up that sample.}
\end{figure}

To present an example we have selected an excited state time correlation
function $C_2(t,t_0)$
of the meson-meson system at relative distance $r=4$. $C_2(t,t_0)$ is the smaller of the
eigenvalues of the $2\times 2$ correlation matrix 
$C_{ij}(t,t_0) = \langle\hat{\Phi}_i^\dagger(t)\hat{\Phi}_j(t_0)\rangle$,
on each time slice.
The reason for selecting this operator is to see how the MEM responds to a data
set that is marginally acceptable, at best.
Figure~\ref{fig15d16d} shows the correlator and the corresponding spectral density obtained
from an average over eight Bayesian fits based on different random annealing start
configurations. The same spectral density is displayed in the first frame of
Fig.~\ref{fig19} on a linear scale.
The dotted lines represent the limits within one standard deviation.
The remaining three frames of Fig.~\ref{fig19} show spectral functions from selected
single start configurations. They illustrate the micro structure fluctuations.

We argue that the micro structure, on a fine discretization scale $\Delta\omega$, is
extraneous information. On the basis that the number of measured data points, as
supplied by the time correlation function $C_v(t,t_0)$, is much smaller than the number of
inferred parameters $\rho_k$, exact knowledge of $\rho$ would actually constitute an
information gain not supported by the data.
Rather, only averages of suitable observables based on the inferred spectral density,
like (\ref{Zn})--(\ref{Dn}) for example, are relevant
information that can be extracted from the Bayesian analysis.
Whether or not the $\rho$ average of a certain observable
is relevant information supported by the data may possibly be decided by the
criterion that the standard deviation with respect to different annealing
starts be small. From Tab.~\ref{tab3} we see that the standard deviations
for the small-moment averages (\ref{Zn},\ref{En},\ref{Dn}) are comparable to
typical gauge configuration statistical errors, for example those in
Tab.~\ref{tab1}. This should be an acceptable test, certainly high resolution
operators would fail it.
\begin{table}
\caption{\label{tab3}Averages of volume, energy, and width of the dominant peak seen in
Figs.~\protect\ref{fig19} over eight random annealing start configurations,
at fixed $\alpha=5.0\times 10^{-7}$ and constant default model $m=10^{-12}$.
The entry $m_{\rm eff,0}$ is the plateau mass (\protect\ref{eff0}) from
Fig.\protect\ref{fig3Rab} with the statistical (jackknife) error,
see Sect.~\protect\ref{sec:plateau}.}
\begin{ruledtabular}
\begin{tabular}{cccc}
 $Z_1$ & $E_1$ & $\Delta_1$ & $m_{\rm eff,0}$\\
\colrule
2156.(11.) & 2.012(7) & 0.214(11) & 1.92(3)
\end{tabular}
\end{ruledtabular}
\end{table}

\subsection{\label{sec:plateau}Relation to plateau methods}

Aside from the obvious differences in algorithm and philosophy
it is important to understand that the traditional plateau method and the
celebrated Bayesian approach also are distinctly different in the way they
utilize the lattice correlator data.
First, the former uses data on only a (subjectively) truncated contiguous set of
time slices while completely ignoring the rest, whereas the latter utilizes
the data on all available time slices without bias.
Second, in the plateau method the stochastic dependence of the data between
the plateau time slices is often ignored\footnote{
Uncorrelated fits to a mass function may be justified if the
number of gauge configurations $N$ is large compared to the
number of plateau times slices $D$, see
\protect\cite{Michael:1994yj,Michael:1995sz}.
There the condition $N>\max(D^2,10(D+1))$ applied to the
situation of Fig.~\protect\ref{fig3} gives $708>169$.}
whereas in the Bayesian approach the dependence is
fully accounted for through the covariance matrix (\ref{Ecov}).
Hence, the traditional plateau method and the Bayesian inference approach
cannot be compared on an equal footing.
In particular, their systematic errors are in principle different. 

\begin{figure}
\includegraphics[angle=0,width=42.0mm]{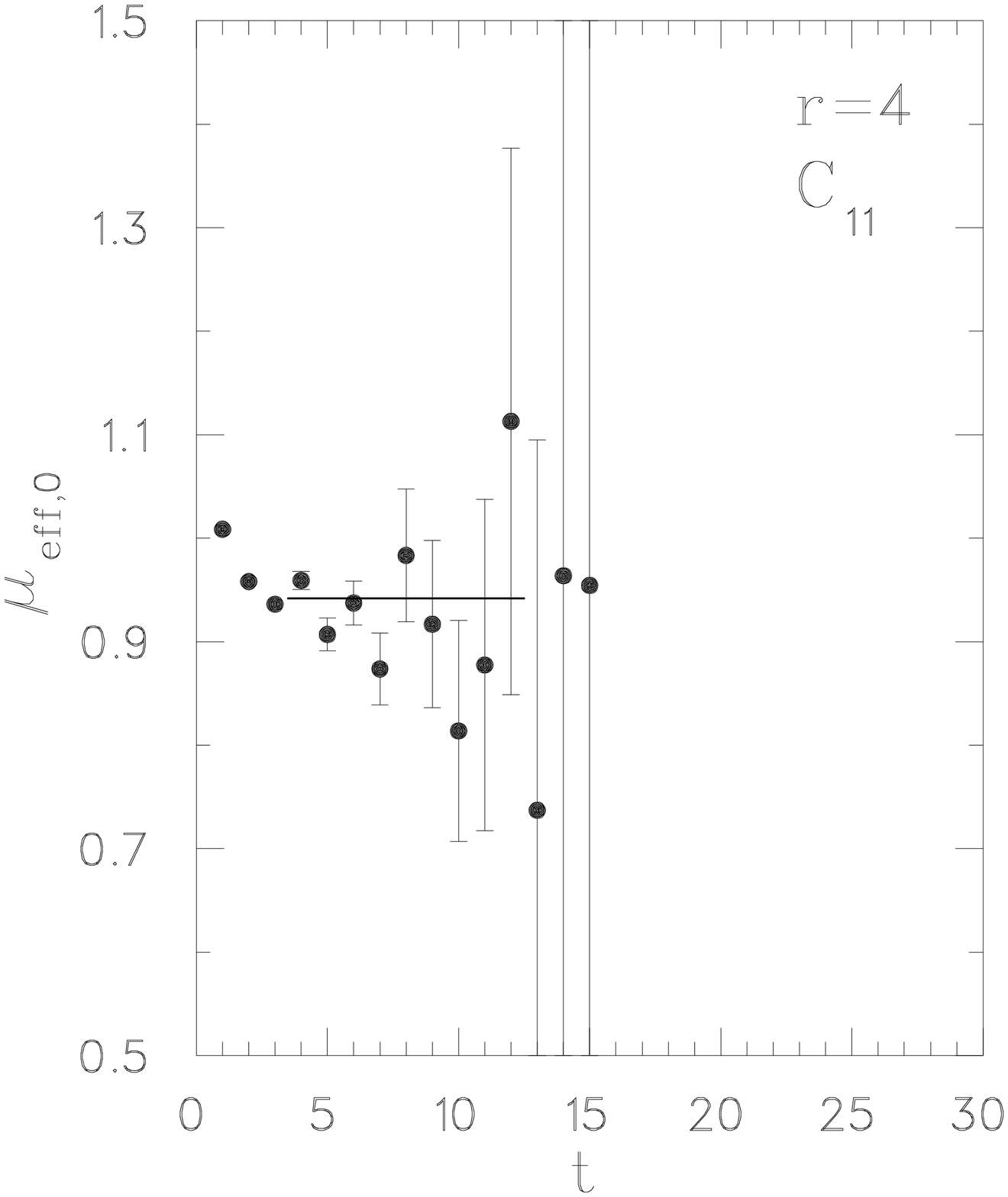}
\includegraphics[angle=0,width=42.0mm]{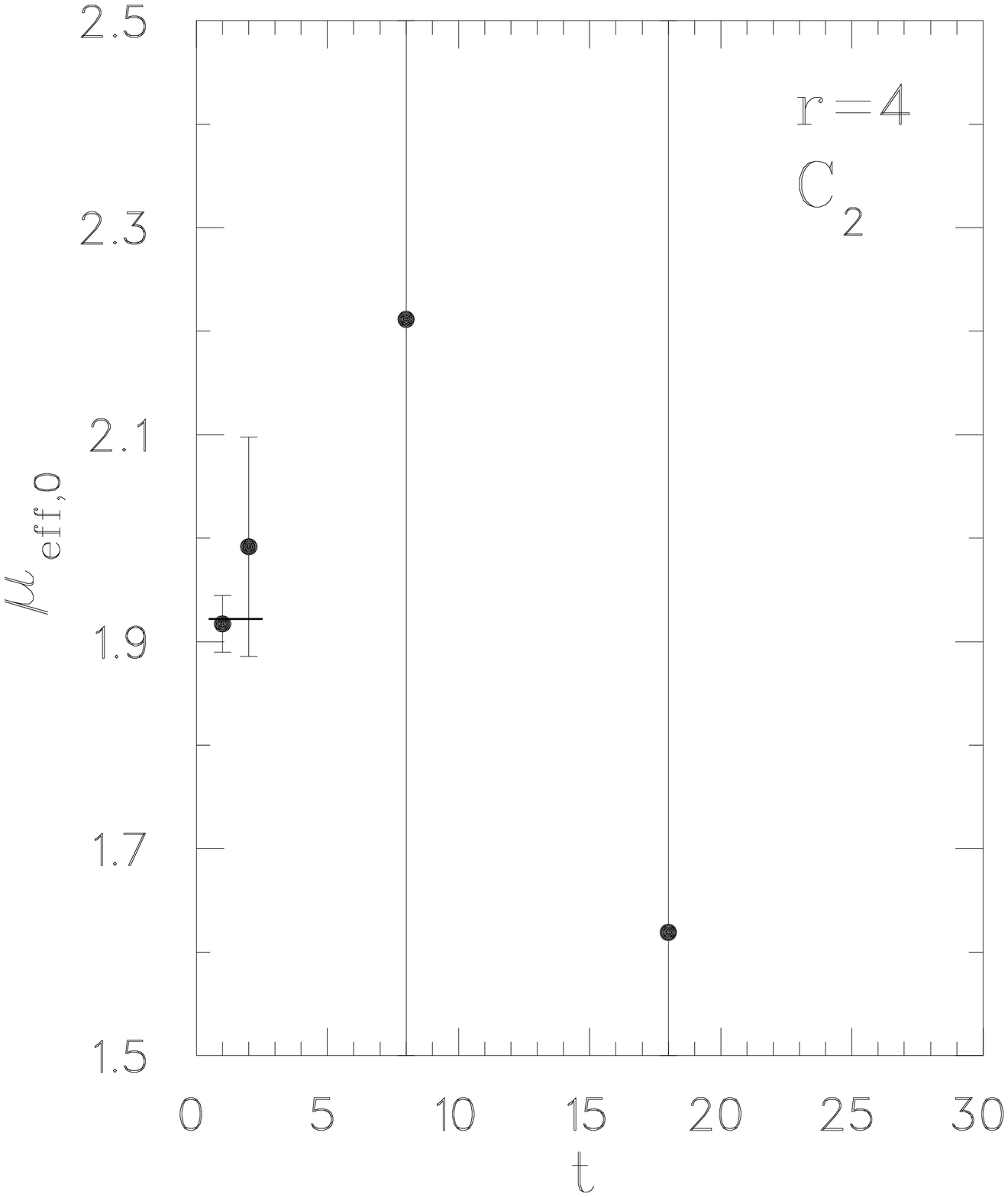}
\caption{\label{fig3Rab}Effective mass functions $\mu_{\rm eff,0}$, see
(\protect\ref{eff0}), of the correlator examples $C_{11}$ and $C_2$ shown
in Figs.\protect\ref{fig6d7d} and \protect\ref{fig15d16d}.
The plateaus are shown as horizontal lines extending over 9 and 2 time
slices, respectively.}
\end{figure}

A comparison of those methods is thus reduced to observing their
responses to the same data sets.
If the numerical quality of data is very good both methods
(in fact any two methods) will of course give the same answers.
An example is the single-meson case discussed above, see Tab.~\ref{tab1}.
In case of imperfect numerical data, however, the two methods
should be expected to give different results.
We illustrate this point by showing in Fig.~\ref{fig3Rab} the effective mass
functions (\ref{eff0}) of the correlators $C_{11}$ and $C_2$ displayed in
Figs.~\ref{fig6d7d} and \ref{fig15d16d}, respectively.
While the $C_{11}$ data are somewhat level within 9 time slices, the
$C_2$ data are extreme in the sense that only 2 data points are available to the
plateau method. Bayesian inference, as illustrated by Fig.~\ref{fig15d16d} and also
Fig.~\ref{fig19}, has no problem responding with a distinct peak. The reason,
of course, is that the entire set of correlator data including their correlations
is available to the Bayesian approach.

In Tabs.~\ref{tab2} and \ref{tab3} we compare the plateau masses $m_{\rm eff,0}$ obtained
from (\ref{eff0}) to the Bayesian results $E_1$. The numbers differ by about 3--5\%. Note
that the statistical (jackknife) errors on the plateau masses are much smaller.
Because of the data truncation the method has no way of `knowing' about the poor quality  
of the correlator data, particularly in the $C_2$ case of the exited state correlator.
The Bayesian method, on the other hand, is fully `aware' of this fact and conveys this
information by responding with a sizable peak width $\Delta_1$, which easily
encompasses the plateau masses.

This raises the question whether Bayesian peak widths or plateau mass
statistical errors are a better measure for the uncertainty of masses extracted
from lattice simulations. The answer is beyond the scope of this work.

\section{\label{sec:conclusion}Summary and conclusion}

We have reported on our experience using Bayesian inference with an
entropic prior, the maximum entropy method, to extract spectral
information from lattice generated time correlation functions.
The latter were taken from a simulation aimed at studying hadronic
interaction, but used here only as a repository of simulation data of diverse quality.

In contrast to other works
the method of choice for extracting spectral densities was simulated annealing.

Between the maximum entropy method and simulated annealing there were three
major concerns about the parameter and algorithm dependence of the results:
Dependence on (i) the entropy weight, (ii) the default model, and (iii) the
annealing start configuration.
Besides suggesting strategies for parameter tuning, independence of the
Bayesian inferred spectral density $\rho$ on (i) the entropy weight, and (ii) the
default model could be demonstrated within a range of eight and fifteen
orders of magnitude of the parameters, respectively. Concerning the annealing
start configuration dependence (iii) we argued that only spectral density
averages of certain operators are acceptable. From an information theory
point of view \cite{Sha49}, those should be operators insensitive to the
micro structure of the inferred spectral density.
In particular, keeping in mind that the theoretical structure of the
lattice spectral function is a superposition of distinct peaks, those operators
include the spectral peak volume $Z_n$, or normalization, the peak energy
$E_n$, or mass, and the peak width $\Delta_n$, or standard deviation.

Bayesian inference has too long been ignored by the lattice community as an
analysis tool. It has an advantage over conventional
plateau methods for extracting hadron masses from lattice simulations
because the entire information contained in the correlator function, or
matrix, is utilized. This aspect is particularly important where
excited state masses are desired, since the noise contamination of their signal
can be significant.
The maximum entropy method is very robust with respect to changing its parameters.
Simulated annealing is practical for obtaining spectral density functions.
The method should be given serious consideration as an alternative for
conventional ways.

\begin{acknowledgments}
This material is based upon work supported by the National Science Foundation
under Grant No. 0073362.
Resources made available through the Lattice Hadron Physics Collaboration (LHPC)
were used in this project.
\end{acknowledgments}



\end{document}